\documentclass[twocolumn,twocolappendix]{aastex63}

\usepackage{listings}
\usepackage{tabularx}
\shorttitle{AGN identification}
\shortauthors{Chang et al.}

\begin{document}

\title{Identifying AGN host galaxies by Machine Learning with HSC+WISE}

\correspondingauthor{Yu-Yen~Chang}
\email{yuyenchang.astro@gmail.com}

\author[0000-0002-6720-8047]{Yu-Yen~Chang}
\affiliation{Department of Physics, National Chung Hsing University, 40227, Taichung, Taiwan}
\affiliation{Academia Sinica Institute of Astronomy and Astrophysics, PO Box 23-141, Taipei 10617, Taiwan}
\author[0000-0001-5615-4904]{Bau-Ching~Hsieh}
\affiliation{Academia Sinica Institute of Astronomy and Astrophysics, PO Box 23-141, Taipei 10617, Taiwan}
\author[0000-0003-2588-1265]{Wei-Hao~Wang}
\affiliation{Academia Sinica Institute of Astronomy and Astrophysics, PO Box 23-141, Taipei 10617, Taiwan}
\author[0000-0001-7146-4687]{Yen-Ting~Lin}
\affiliation{Academia Sinica Institute of Astronomy and Astrophysics, PO Box 23-141, Taipei 10617, Taiwan}
\author[0000-0003-1213-9360]{Chen-Fatt~Lim}
\affiliation{Graduate Institute of Astrophysics, National Taiwan University, Taipei 10617, Taiwan}
\affiliation{Academia Sinica Institute of Astronomy and Astrophysics, PO Box 23-141, Taipei 10617, Taiwan}
\author[0000-0002-3531-7863]{Yoshiki~Toba}
\affiliation{Department of Astronomy, Kyoto University, Kitashirakawa-Oiwake-cho, Sakyo-ku, Kyoto 606-8502, Japan}
\affiliation{Academia Sinica Institute of Astronomy and Astrophysics, PO Box 23-141, Taipei 10617, Taiwan}
\affiliation{Research Center for Space and Cosmic Evolution, Ehime University, 2-5 Bunkyo-cho, Matsuyama, Ehime 790-8577, Japan}
\author{Yuxing~Zhong}
\affiliation{Department of Physics, Waseda University, 1-6-1 Nishiwaseda Shinjuku-ku, Tokyo 169-8050, Japan}
\author{Siou-Yu~Chang}
\affiliation{Department of Physics, National Chung Hsing University, 40227, Taichung, Taiwan}



\begin{abstract}
We use machine learning techniques to investigate their performance in classifying active galactic nuclei (AGNs), including X-ray selected AGNs (XAGNs), infrared selected AGNs (IRAGNs), and radio selected AGNs (RAGNs). Using known physical parameters in the Cosmic Evolution Survey (COSMOS) field, we are able to  well-established training samples in the region of Hyper Suprime-Cam (HSC) survey. We compare several Python packages (e.g., \texttt{scikit-learn}, \texttt{Keras}, and \texttt{XGBoost}), and use \texttt{XGBoost} to identify AGNs and show the performance (e.g., accuracy, precision, recall, F1 score, and AUROC). Our results indicate that the performance is high for bright XAGN and IRAGN host galaxies. The combination of the HSC (optical) information with the Wide-field Infrared Survey Explorer (WISE) band-1 and WISE band-2 (near-infrared) information perform well to identify AGN hosts. For both type-1 (broad-line) XAGNs and type-1 (unobscured) IRAGNs, the performance is very good by using optical to infrared information. These results can apply to the five-band data from the wide regions of the HSC survey, and future all-sky surveys. 
\end{abstract}



\section{Introduction} \label{sec1}
The processes driving the co-evolution of galaxies, active galactic nuclei (AGNs),  and their super-massive black holes remain a largely debated issue in extragalactic astrophysics. 
Specifically, the connection between the formation of stars in galaxies and the fueling of their central black holes is still not fully understood.  Different techniques have been investigated to select and identify various AGN sample \citep[e.g., ][and reference therein]{1981PASP...93....5B, 1995PASP..107..803U, 2003MNRAS.346.1055K, 2003MNRAS.343..585F, 2004ApJS..154..166L, 2005ApJ...631..163S, 2006ApJ...640..167A, 2006MNRAS.372..961K, 2006A&A...451..457T, 2008ARA&A..46..475H, 2011ApJ...728...58B, 2011ApJ...736..104J, 2012ARA&A..50..455F, 2017A&ARv..25....2P, 2018ARA&A..56..625H}.  However, it is still difficult to properly select AGNs and avoid sample selection bias. 

The increasing astronomical data has led to a need for machine learning (ML) methods. For examples, decision trees \citep[e.g.,][]{2006ApJ...650..497B,2011AJ....141..189V,2015A&C....11...64S}, support vector machines \citep[e.g.][]{2015MNRAS.448.1305K}, neural networks (NN) \citep[e.g.,][]{1992AJ....103..318O,1996A&AS..117..393B}, convolutional neural networks (CNN) \citep[e.g.,][]{2015MNRAS.450.1441D,2015ApJS..221....8H,2016A&A...596A..39K,2017MNRAS.464.4463K,2018MNRAS.476.3661D}, and XGBoost \citep[e.g.,][]{2019MNRAS.489.1770L,2019MNRAS.490.2367C,2020ApJ...895..104L, 2021MNRAS.503.4136G} have been used to deal with data in astronomy and astrophysics.

Recently, ML has been applied to derive various physical parameters of galaxies \citep[e.g.,][]{2015ApJ...813...53M,2016A&A...596A..39K,2018A&A...609A.111D,2019ApJ...881L..14H,2019MNRAS.489.4817D,2019A&A...622A.137B}. In particular, classification is one of the important issues on galaxy and AGN properties \citep[e.g.,][and references therein]{2017A&ARv..25....2P,2018ARA&A..56..625H}. Therefore, ML can be used to classify AGNs efficiently \citep[e.g.,][]{2008MNRAS.391..369C,2014ApJ...782...41D,2011ApJ...735...68K,2014MNRAS.437..968C,2016ApJ...820....8S,2018A&A...619A..14F,2019AJ....157....9B,2019ApJ...881L...9F, 2021MNRAS.501.3951C}, and will be helpful for us to identify different kinds of AGN hosts from non-AGN galaxies. 

In this paper, we will compare several algorithms from \texttt{scikit-learn}, \texttt{Keras}, and \texttt{XGBoost}, and show our results by using the state-of-the-art ML methods from \texttt{XGBoost} to identify X-ray selected AGN (XAGN), infrared (IR) selected AGN (IRAGN), and radio selected AGN (RAGN) host galaxies with Hyper Suprime-Cam (HSC) and Wide-field Infrared Survey Explorer (WISE) data.
The structure of this paper is as follows.
We describe the data and sample selections in Section \ref{sec2}.
We analyze the properties in Section \ref{sec3}.
We discuss the results in Section \ref{sec4} and summarize in Section \ref{sec5}.
Throughout the paper, we use AB magnitudes, adopt the cosmological parameters
($\Omega_{\rm M}$,$\Omega_\Lambda$,$h$)=(0.30,0.70,0.70), and assume the
stellar initial mass function of \citet{2003PASP..115..763C}.

\begin{figure}
\centering
\includegraphics[width=0.99\columnwidth]{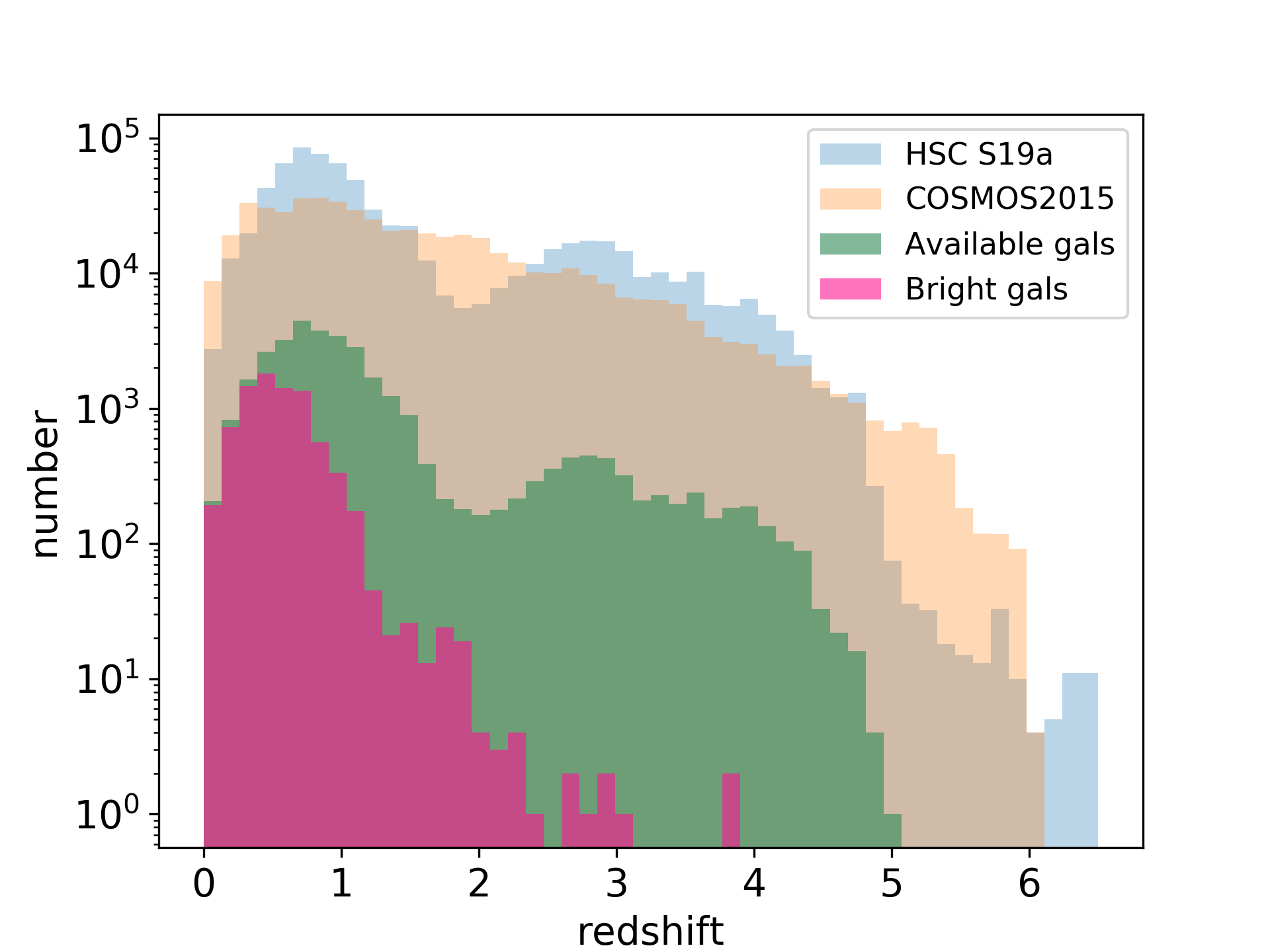} 
\caption{Redshift distribution for the HSC-SSP S19a, COSMOS2015, our 32,204 available galaxies, and 8,193 bright galaxies.}
\label{redshift}
\end{figure}

\begin{figure}
\centering
\includegraphics[width=0.99\columnwidth]{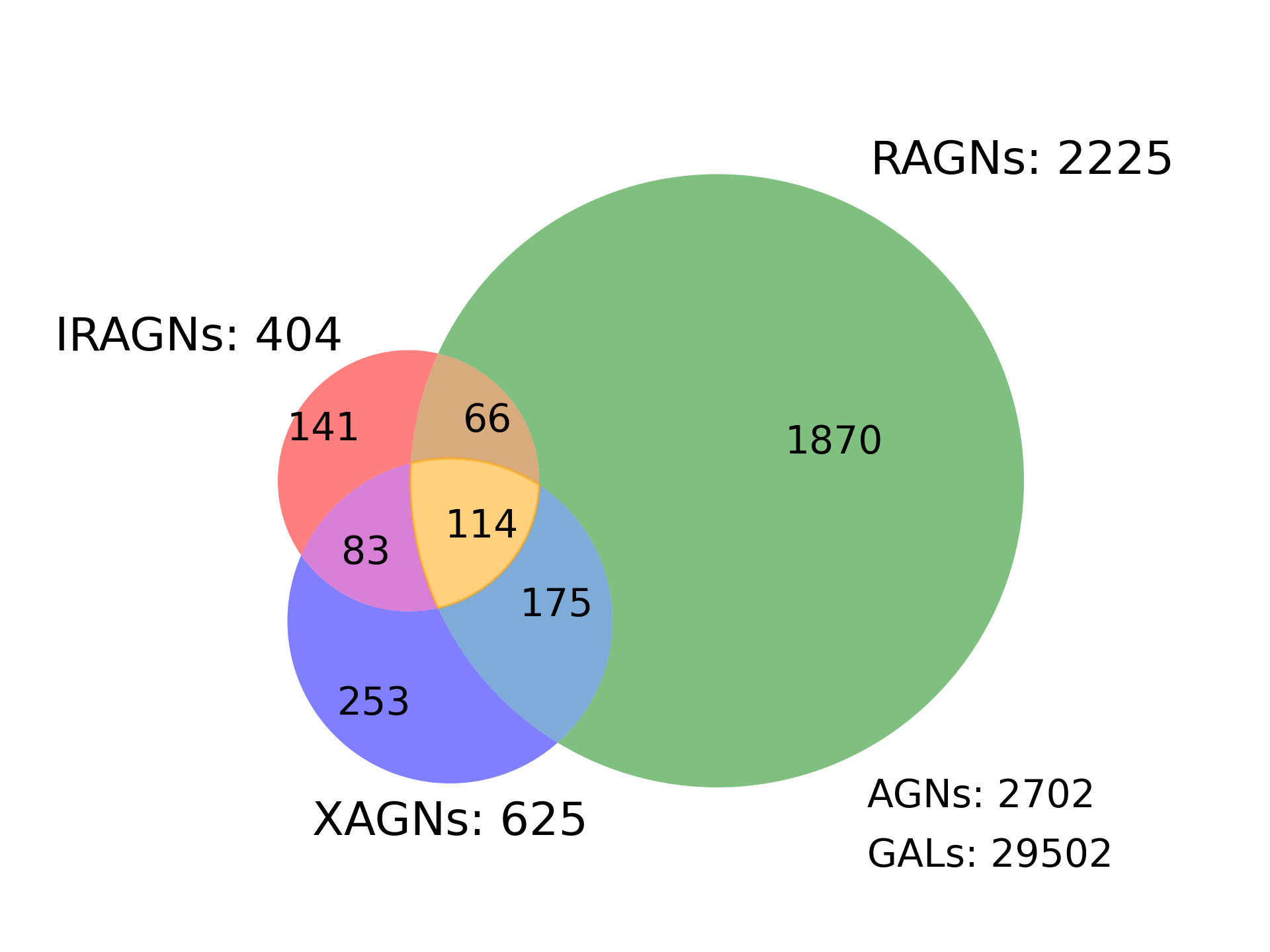} 
\caption{Among the 32,204 objects, there are 2,702 AGNs (including 625 XAGNs, 404 IRAGNs, and 2,225 RAGNs) as well as 29,502 GALs.}
\label{ga}
\end{figure}

\begin{figure}
\centering
\includegraphics[width=0.99\columnwidth]{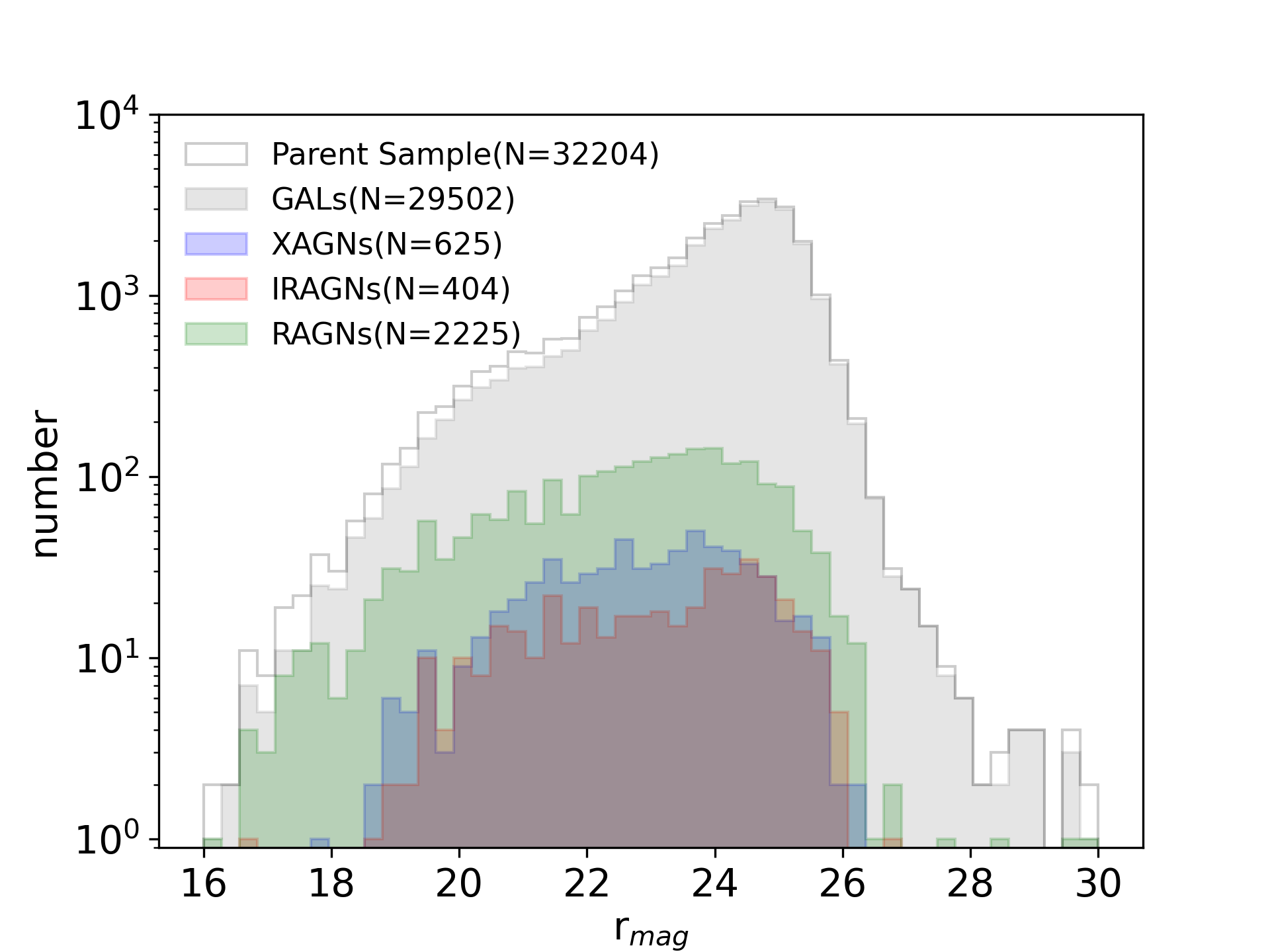} 
\caption{Magnitude distributions of the 625 XAGNs, the 404 IRAGNs, the 2,225 AGNs, the 29,502 GALs galaxies, and the whole 32,204 parent sample.}
\label{mag}
\end{figure}

\section{Data} \label{sec2}
We use the Wide layer of optical photometry from the Hyper Suprime-Cam \citep{2012SPIE.8446E..0ZM} Subaru Strategic Program\footnote{http://hsc.mtk.nao.ac.jp/} (HSC-SSP). 
The HSC-SSP is an optical imaging survey with five broadband filters ($grizy$-band) and four narrow-band filters \citep[see][]{2018PASJ...70S...4A,2018PASJ...70S...5B}.
HSC-SSP consists of three layers: Wide, Deep, and UltraDeep. This work uses S19a Wide-layer data \citep[s19a\_wide.forced; e.g.,][]{2018PASJ...70S...1M,2018PASJ...70S...4A,2019PASJ...71..114A,2019ApJS..243...15T}. The HSC-SSP Wide-layer covers six fields (XMM-LSS, GAMA09H, WIDE12H, GAMA15H, HECTOMAP, and VVDS). The typical seeing is about 0$\arcsec$.6 in the $i$-band and the astrometric uncertainty is about 40 milliarcsecond in rms. There are 2,802,212 objects in the S19a Wide (HSC-Wide) catalog. Among them, 705,161 objects have photometric redshift (photoz\_best in s19a\_wide.photoz\_mizuki) information, and 62,567 objects have spectroscopic redshift (specz\_redshift in s19a\_wide.specz) information.

We match the HSC-Wide catalog with the 1,182,108 objects in the COSMOS 2015 catalog \citep{2016ApJS..224...24L} within 1$\arcsec$, and 660,352 objects are remaining. The spectral energy distributions (SEDs) fitting properties are from \citet{2017ApJS..233...19C}, which derived by the MAGPHYS+AGN fitting technique \citep{2008MNRAS.388.1595D,2015ApJ...806..110D}, in the Cosmic Evolution Survey \citep[COSMOS;][]{2007ApJS..172....1S} field. 

We include infrared information by using ALLWISE data \citep{2010AJ....140.1868W,2011ApJ...731...53M} at 3.4, 4.6, 12, and 22 $\mu$m (W1, W2, W3, and W4), identified as the brightest object within a search radius of 6 arcsec, similar to the WISE PSF and as the SDSS+WISE sample in \citet{2015ApJS..219....8C}. In order to avoid missing value problem, we focus on 32,204 galaxies with available photometric redshifts \citep[e.g., ][we adopted MIZUKI photometric redshift in this paper]{2014ApJ...792..102H,2015ApJ...801...20T}, as well as detections in HSC $grizy$ bands, W1, and W2. To reach better performances, we also investigated a subsample of 8,193 bright galaxies with $R_{mag}<$23 (see Subsection 3.3 for more details). The redshift distribution is shown in Figure~\ref{redshift}. 

The matched catalog contains optical information from HSC-Wide and infrared information from ALLWISE in the COSMOS field, which provides the training and test sets in our sample. 
We separate the sample into XAGNs (625 XAGNs: $\rm{L}_X$(2-10 keV)$>10^{42}$ ergs/s, absorption-corrected), IRAGNs (404 IRAGNs: using the mid-infrared colors from the IRAC 3.6, 5.8, 4.5, and 8.0 $\mu$m as defined in \citet{2017ApJS..233...19C}), RAGNs (2,225 RAGNs:  $>$ 3-$\sigma$ radio excess in $\log(L_{1.4GHz}/SFR_{IR}$ as defined in \citet{2017A&A...602A...6S,2017A&A...602A...3D}), and non-AGN galaxies (29,502 GALs) that are identified by previous work \citep{2016ApJ...819...62C,2016ApJ...817...34M,2017ApJS..233...19C,2017MNRAS.466L.103C,2017A&A...602A...6S,2017A&A...602A...3D}. As shown in Figure~\ref{ga}, the total number of AGNs is 2,702, and some AGNs are belong to two or three AGN-selection methods. Magnitude distributions of the 625 XAGNs, the 404 IRAGNs, the 2,225 AGNs, the 29,502 non-AGN galaxies, and the 32,204 parent sample are shown in Figure~\ref{mag}.

In this paper, photometric redshifts from HSC-Wide are adopted for the 32,204 parent sample, including the 2,702 AGNs.  We check the photometric redsfhit quality of our sample by calculating the catastrophic error ($\eta=|z_2-z_1|/(1+z_1) >0.15$) and the redshift accuracy ($\sigma_{\Delta z/(1+z_1)}$=$\sigma_{|z_2-z_1|/(1+z_1)}$) similar to \citet{2009ApJ...690.1236I}, where $z_2$ is the photometric redshift from the HSC-Wide catalog and $z_1$ is the redshift from compared sample. Comparing to available photometric redshift in the COSMOS 2015 catalog, the catastrophic error and redshift accuracy in the AGN sample (625 XAGN: $\eta$=30.1\% and $\sigma_{\Delta z/(1+z_1)}$=0.0847; 400 IRAGN: $\eta$=39.2\% and $\sigma_{\Delta z/(1+z_1)}$=0.1082; 1,942 RAGN: $\eta$=15.4\% and $\sigma_{\Delta z/(1+z_1)}$=0.0660) is not small but not far from the whole parent sample (22,721 objects: $\eta$=23.8\% and $\sigma_{\Delta z/(1+z_1)}$=0.0744).  
The photometric redshift errors show dependence on magnitude, so both catastrophic error and redshift accuracy can be a function of depth. For bright sources ($r_{mag}<$23), the catastrophic error and the redshift accuracy in the AGN sample (312 XAGN: $\eta$=22.4\% and $\sigma_{\Delta z/(1+z_1)}$=0.0847; 172 IRAGN: $\eta$=29.5\% and $\sigma_{\Delta z/(1+z_1)}$=0.08; 1,024 RAGN: $\eta$=7.3\% and $\sigma_{\Delta z/(1+z_1)}$=0.0474) and the parent sample (6,148 objects: $\eta$=7.8\% and $\sigma_{\Delta z/(1+z_1)}$=0.0432) are smaller. Besides, photometric redshift from COSMOS have themselves errors associated \citep[e.g., ][]{2019NatAs...3..212S}, so it is possible to have some redshift differences between the HSC-Wide and the COSMOS catalogs. 
Comparing to available spectroscopic redshift in HSC-Wide, the catastrophic error and redshift accuracy in the AGN sample (442 XAGN: $\eta$=4.8\% and $\sigma_{\Delta z/(1+z_1)}$=0.0005; 227 IRAGN: $\eta$=7.5\% and $\sigma_{\Delta z/(1+z_1)}$=0.0005; 1,254 RAGN: $\eta$=2.7\% and $\sigma_{\Delta z/(1+z_1)}$=0.0005) is also close to the whole parent sample (7,701 objects: $\eta$=4.1\% and $\sigma_{\Delta z/(1+z_1)}$=0.0005). 
A possible explanation for the better results with the comparison to available spectroscopic redshift can be that most of the matched sources are bright (60\% of them are $r_{mag}<$23 sources and 97\% of them are $r_{mag}<$25 sources). As mentioned earlier, brighter sources show smaller photometric redshift errors.


\begin{figure*}
\centering
\includegraphics[width=0.32\textwidth]{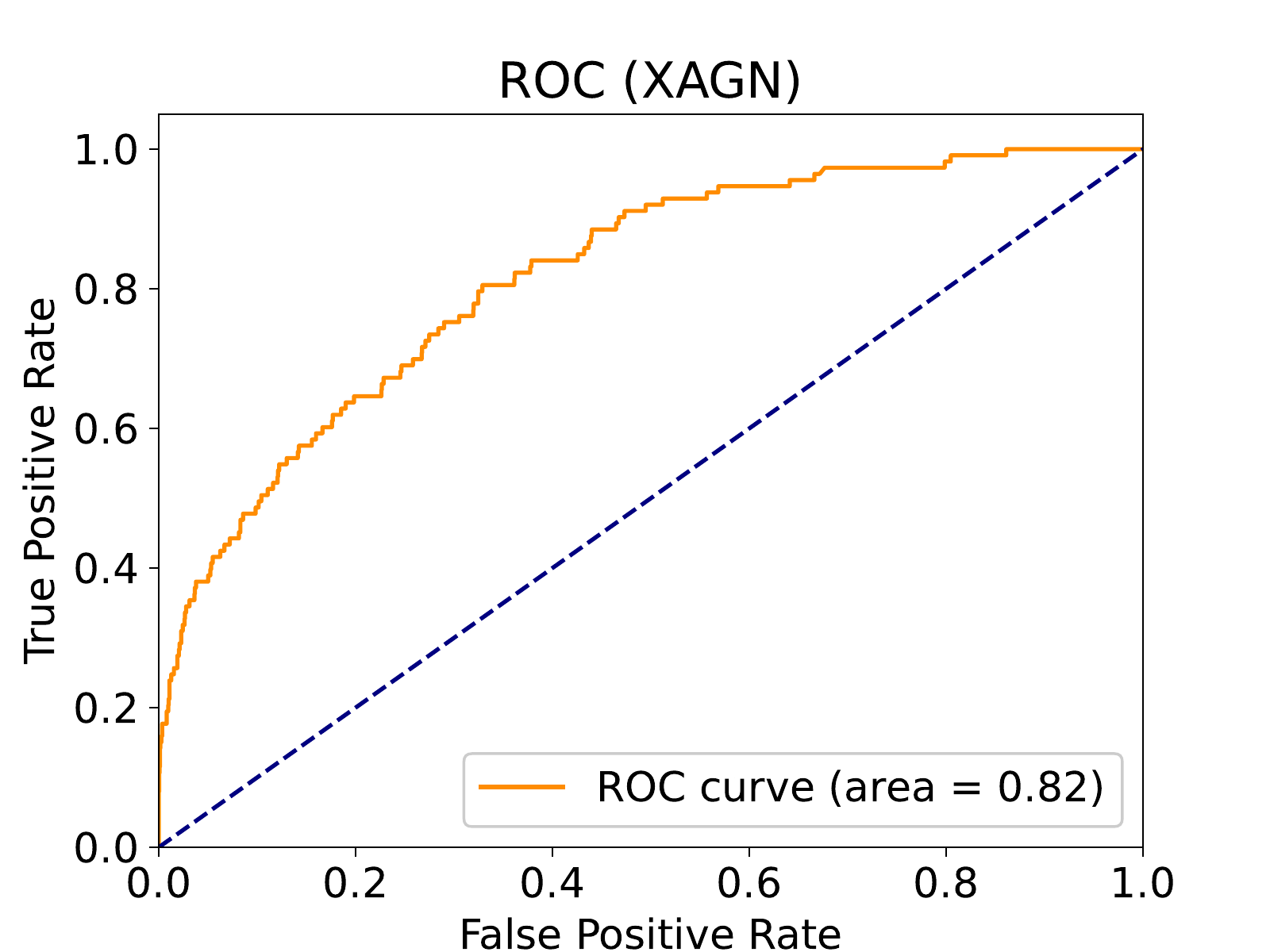} 
\includegraphics[width=0.32\textwidth]{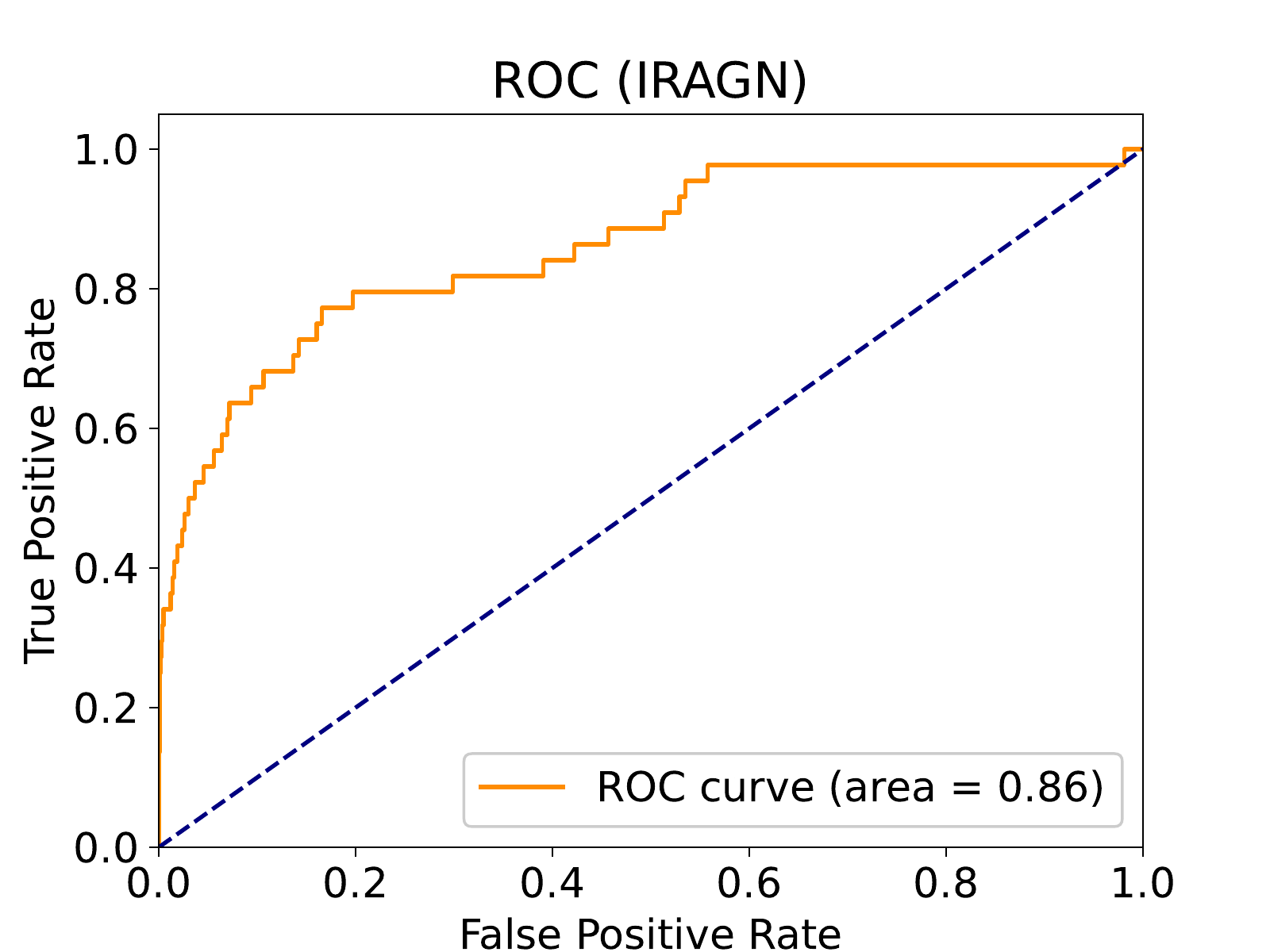} 
\includegraphics[width=0.32\textwidth]{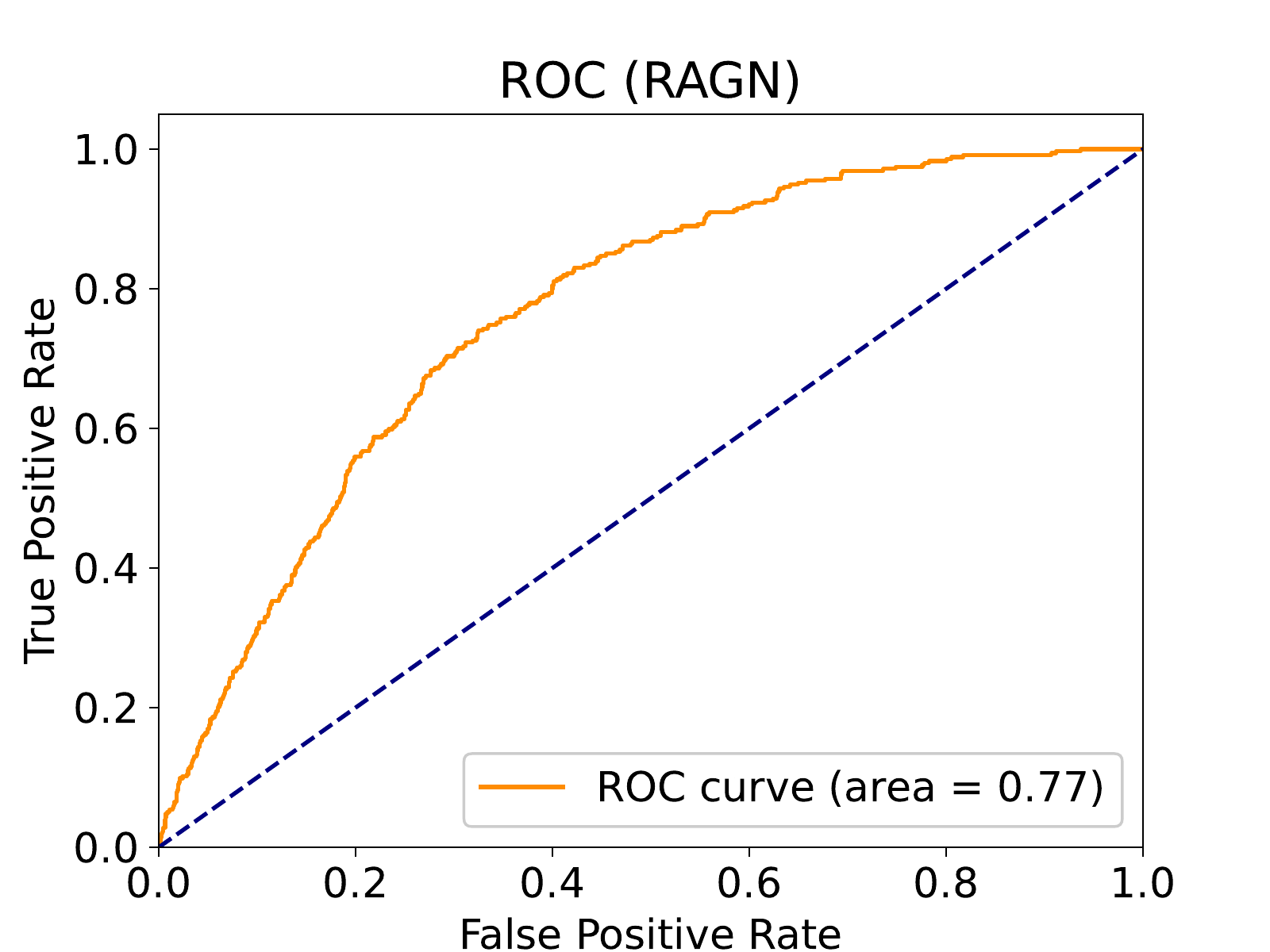} 
\caption{ROC (Receiver Operating Characteristic) curve example for XAGNs, IRAGNs, and RAGNs ($r_{mag}<$23; N=8,193).}
\label{xgb_roc}
\end{figure*}

\section{Analysis} \label{sec3}

\subsection{Evaluations}

We evaluate the quality of the classification schemes with their performance. 
First, we defined true positive (TP; an AGN source which is classified as AGN), true negative (TN: a non-AGN source which is not classified as AGN), false positive (FP: a non-AGN source which is classified as AGN), and false negative (FN: an AGN source which is not classified as AGN). 
Therefore, the true positive rate (TPR), the true negative rate (TNR), the false positive rate (FPR), and the false negative rate (FNR) are $TPR = TP/(TP+FN)$, $TNR = TN/(TN+FP)$, $FPR = FP/(TN+FP)$, and $FNR = FN/(TP+FN)$, respectively. A good classification will classify an AGN as an AGN (high TPR) and a non-AGN as a non-AGN (high TNR), rather than a non-AGN as an AGN (low FPR) and an AGN as a non-AGN (low FNR). Therefore, high performance can be determined by high accuracy, high precision, high recall, high F1 score, and high Area Under the Receiver Operating Characteristic (AUROC) as described below. 

\begin{itemize}
\item Accuracy: fraction of sources (AGN and non-AGN) which are classified correctly over all sources. 
\begin{equation}
ACC=\frac{TP+TN}{TP+TN+FP+FN}
\end{equation}
\item Precision: AGN sources which are classified correctly as AGNs over all classified AGNs. 
\begin{equation}
P=\frac{TP}{TP+FP}
\end{equation}
\item Recall: AGN sources which are classified correctly as AGNs over all AGN sources. 
\begin{equation}
R=\frac{TP}{TP+FN}
\end{equation}
\item F1 score:  a harmonic mean of the precision and the recall. 
\begin{equation}
F1=2\times\frac{P\times R}{P+R}
\end{equation}
\item AUROC: Area Under the Receiver Operating Characteristic (ROC) curve as shown in the ROC curve (TP versus FP) of Figure~\ref{xgb_roc}.
\end{itemize}

\subsection{Techniques and Parameters}

We use several algorithms from Python packages, \texttt{scikit-learn} \footnote{http://scikit-learn.org/}, \texttt{Keras} \footnote{https://keras.io/}, and \texttt{XGBoost}\footnote{https://xgboost.readthedocs.io/} to identify XAGNs, IRAGNs, RAGNs, and GALs.  
 
First, we use the logistic regression and the random forest classifier algorithms in \texttt{scikit-learn}.  The logistic regression provides basic logistic function to model a binary dependent variable. The performance (e.g., accuracy, precision, recall, F1 score, and AUROC) have no significant changes after about 100 iterations. We choose 1000 as the maximum number of iterations (max\_iter=1000) in Table~\ref{tab_tfpnr}.

The random forest classifier in \texttt{scikit-learn} provides an estimator which fits a number of decision tree classifiers on various sub-samples of the data and uses averaging to improve the performance and control over-fitting. The performance have no significant changes after about 100 number of trees. We choose 1000 as the number of trees in the forest (n\_estimators=1000)  in Table~\ref{tab_tfpnr}.

We use the sequential model in \texttt{Keras}, which is a deep learning application programming interface written in Python, running on top of the machine learning platform \texttt{TensorFlow}. A Sequential model can stack pain layers where each layer has one input tensor and one output tensor. By considering the computing speed and the performance, we use early stopping (patience = 5; stop while lack of improvement after 5 epochs) with the callback function and 1000 epochs of model fit. As a result, all the training stop before 100 epochs of model fit in Table~\ref{tab_tfpnr}. We use Flatten, Dense, Dropout layers for this work. The setting of 2 in the last Dense layer set the whole sequential model as a binary classifier.\\
\begin{lstlisting}[language=Python]
model = Sequential()
model.add(Convolution1D(32, 3, 
	border_mode = 'same', 
	input_shape = (inp, 1)))
model.add(Convolution1D(32, 3, 
	border_mode='same'))
model.add(Flatten())
model.add(Dense(128, activation = 'relu'))
model.add(Dropout(0.5))
model.add(Dense(2, activation='softmax'))
\end{lstlisting}

We also use \texttt{XGBoost}, which is a machine learning algorithms to optimize distributed gradient boosting library and provides a parallel tree boosting learning. We use early stopping (early\_stopping\_rounds = 5; stop while lack of improvement after 5 rounds) and 1000 number of round of iteration. As a result, all the training stop before 300 number round of iteration in Table~\ref{tab_tfpnr}. We choose the following parameters by considering the computing speed and the performance with grid-searching technique.\\
\begin{lstlisting}[language=Python]
param = {
	'max_depth': 5,                
	'eta': 0.3,                                 
	'objective': 'multi:softprob', 
	'num_class': 2}                
\end{lstlisting}

For the above algorithms, we split our data to training (67\%) and test (33\%) samples. We test the algorithms to reach their best performance by choosing the input parameters as mentioned above.  Because AGNs only represent a small fraction ($<$10\%) of the whole sample, we randomly chose the same number of the selected galaxies for the selected AGNs in the training sample. In other words, we used over-sampling technique to deal with the imbalance problem. 		

\begin{table*}[t]
\centering
\begin{tabular}{cccccccccc}
\hline
\hline
type & TNR & FPR & FNR & TPR & ACC & P & R & F1 & AUROC  \\
\hline
\hline
\multicolumn{10}{c}{Logistic Regression: HSC+W12} \\ 
\hline
XAGN & 0.76$\pm$0.03 & 0.24$\pm$0.03 & 0.39$\pm$0.05 & 0.61$\pm$0.05 & 0.78$\pm$0.03 & 0.52$\pm$0.00 & 0.68$\pm$0.02 & 0.48$\pm$0.01 & 0.68$\pm$0.02 \\
IRAGN & 0.55$\pm$0.07 & 0.45$\pm$0.07 & 0.33$\pm$0.08 & 0.67$\pm$0.08 & 0.54$\pm$0.05 & 0.51$\pm$0.00 & 0.62$\pm$0.02 & 0.39$\pm$0.02 & 0.62$\pm$0.02 \\
RAGN & 0.79$\pm$0.02 & 0.21$\pm$0.02 & 0.39$\pm$0.03 & 0.61$\pm$0.03 & 0.78$\pm$0.01 & 0.58$\pm$0.00 & 0.70$\pm$0.01 & 0.58$\pm$0.01 & 0.70$\pm$0.01 \\
\hline
\multicolumn{10}{c}{Random Forest: HSC+W12} \\ 
\hline
XAGN & 1.00$\pm$0.00 & 0.00$\pm$0.00 & 0.89$\pm$0.04 & 0.11$\pm$0.04 &  0.98$\pm$0.00 & 0.93$\pm$0.01 & 0.56$\pm$0.02 & 0.60$\pm$0.02 & 0.56$\pm$0.02 \\
IRAGN & 1.00$\pm$0.00 & 0.00$\pm$0.00 & 0.90$\pm$0.05 & 0.10$\pm$0.05 & 0.99$\pm$0.00 & 0.83$\pm$0.01 & 0.55$\pm$0.02 & 0.84$\pm$0.02 & 0.55$\pm$0.02 \\
RAGN & 0.58$\pm$0.01 & 0.00$\pm$0.00 & 0.90$\pm$0.02 & 0.10$\pm$0.02 & 0.92$\pm$0.00 & 0.68$\pm$0.01 & 0.54$\pm$0.01 & 0.55$\pm$0.01 & 0.55$\pm$0.01 \\
\hline
\multicolumn{10}{c}{Keras: HSC+W12} \\ 
\hline
XAGN & 0.98$\pm$0.03 & 0.02$\pm$0.03 & 0.86$\pm$0.09 & 0.14$\pm$0.09 & 0.94$\pm$0.02 & 0.53$\pm$0.06 & 0.59$\pm$0.04 & 0.54$\pm$0.02 & 0.78$\pm$0.02 \\
IRAGN & 0.99$\pm$0.01 & 0.01$\pm$0.01 & 0.86$\pm$0.04 & 0.14$\pm$0.04 & 0.98$\pm$0.01 & 0.57$\pm$0.03 & 0.57$\pm$0.02 & 0.57$\pm$0.02 & 0.65$\pm$0.03 \\
RAGN & 0.86$\pm$0.03 & 0.14$\pm$0.03 & 0.53$\pm$0.08 & 0.47$\pm$0.08 & 0.84$\pm$0.03 & 0.58$\pm$0.01 & 0.60$\pm$0.03 & 0.60$\pm$0.01 & 0.80$\pm$0.01 \\
\hline
\multicolumn{10}{c}{XGBoost: HSC+W12} \\ 
\hline
XAGN & 0.98$\pm$0.03 & 0.02$\pm$0.03 & 0.75$\pm$0.05 & 0.25$\pm$0.05  & 0.97$\pm$0.05 & 0.60$\pm$0.08 & 0.61$\pm$0.02 & 0.60$\pm$0.08 & 0.80$\pm$0.02 \\
IRAGN & 1.00$\pm$0.01 & 0.01$\pm$0.02 & 0.78$\pm$0.05 & 0.22$\pm$0.05  & 0.99$\pm$0.02 & 0.64$\pm$0.08 & 0.62$\pm$0.03 & 0.62$\pm$0.06 & 0.72$\pm$0.03 \\
RAGN & 0.84$\pm$0.03 & 0.16$\pm$0.03 & 0.40$\pm$0.04 & 0.60$\pm$0.04 & 0.88$\pm$0.01 & 0.61$\pm$0.03 & 0.63$\pm$0.01 & 0.63$\pm$0.02 & 0.83$\pm$0.01 \\
\hline
\hline
\end{tabular}
\caption{Numbers of TNR, FPR, FNR, TPR,  accuracy, precision, recall, F1 score, and AUROC of different algorithms (logistic regression, random forest, \texttt{Keras}, and \texttt{XGBoost}) for XAGNs, IRAGNs, and RAGNs (All available objects; N=32,204). The uncertainties are derived by bootstrapping.}
\label{tab_tfpnr}
\end{table*} 

\begin{figure}
\centering
\includegraphics[width=0.99\columnwidth]{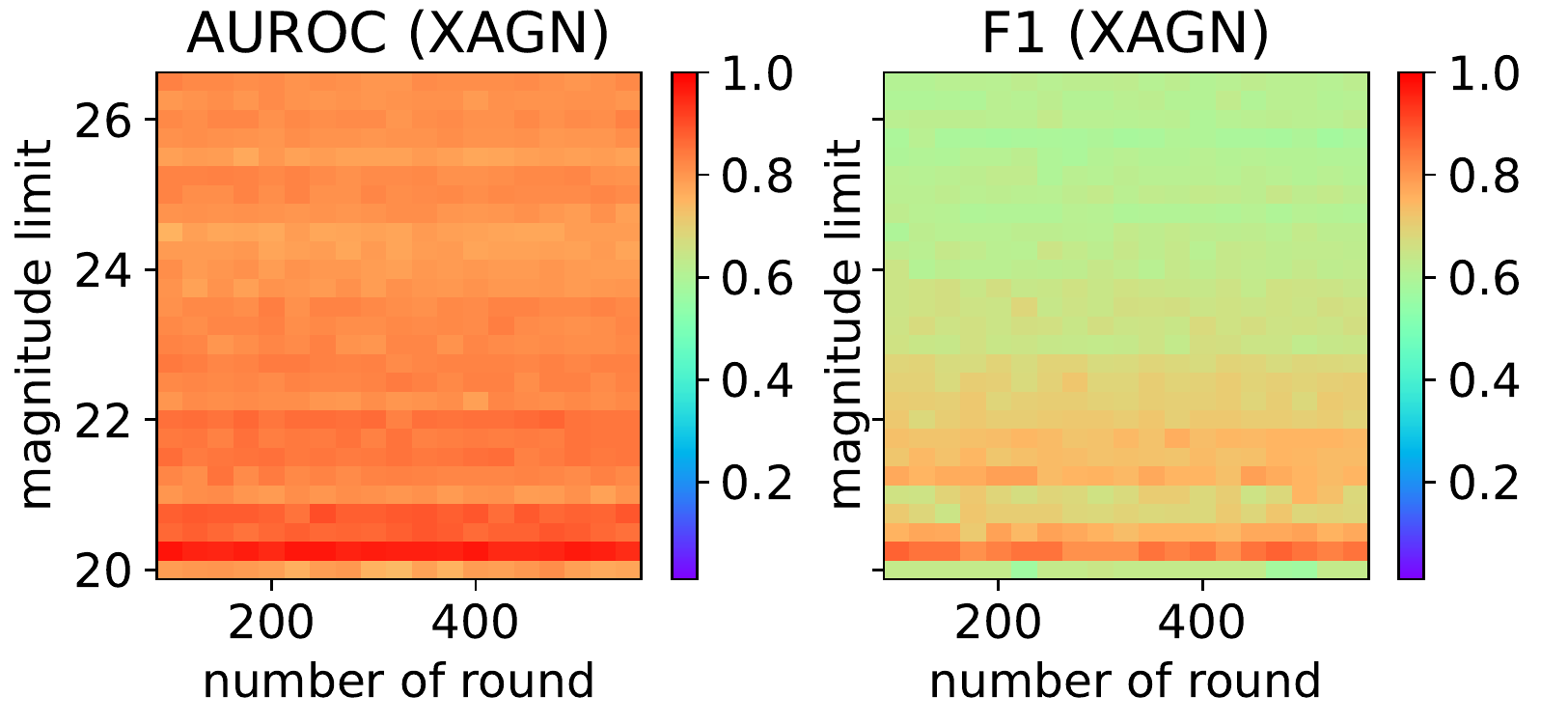} 
\includegraphics[width=0.99\columnwidth]{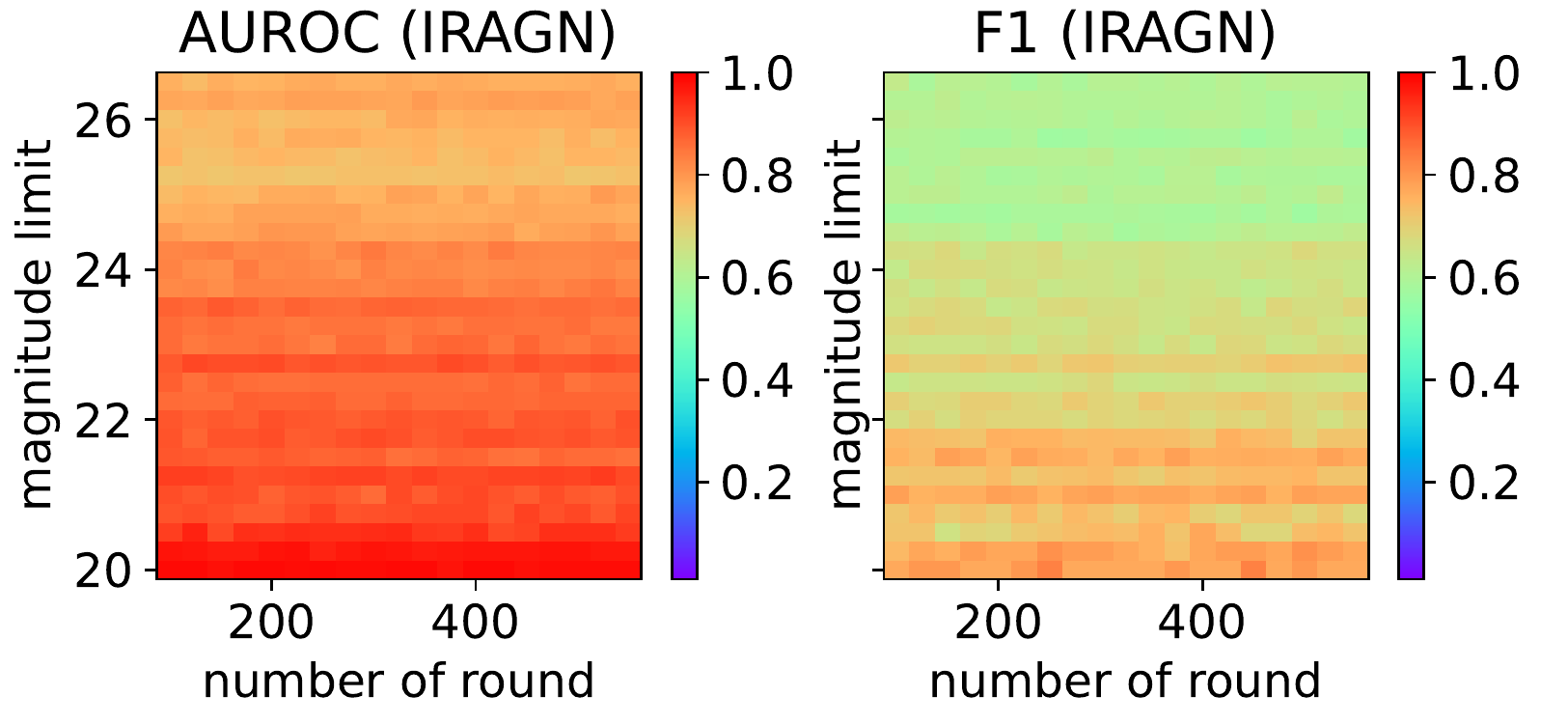} 
\includegraphics[width=0.99\columnwidth]{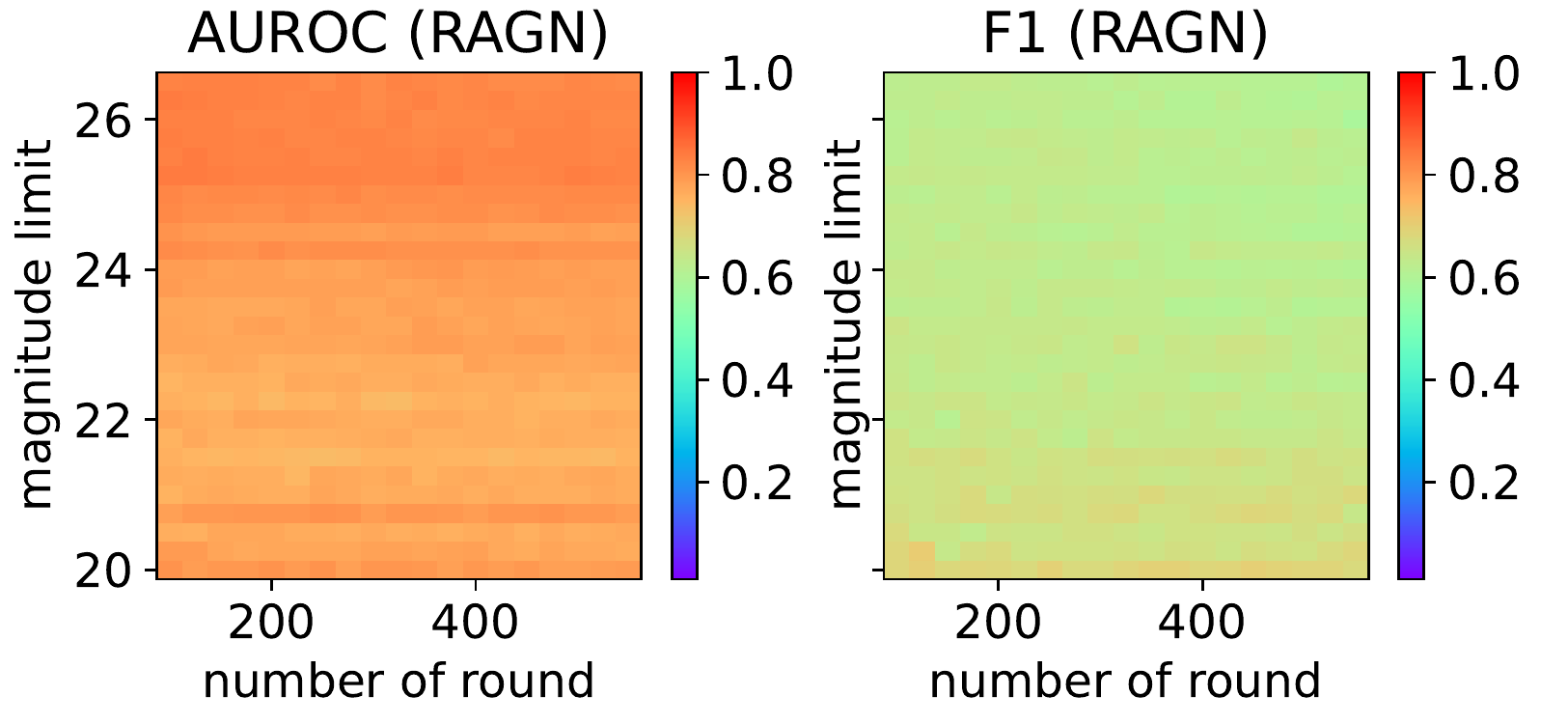} 
\caption{AUROC and F1 score grid-searching example for all available objects (N=32,204) in \texttt{XGBoost}. The x-axis is the number of round in \texttt{XGBoost}, and the y-axis is the $r$-band magnitude limit. The color coding shows ROC and F1 score, respectively. For XAGNs and IRAGNs, lower $r$-band magnitude limits (e.g., $r_{mag}<$23) perform better.}
\label{xgb_roc_f1}
\end{figure}

\subsection{Performances}\label{sec33}
In Table~\ref{tab_tfpnr}, we compare TNR, FPR, FNR, TPR,  accuracy, precision, recall, F1 score, and AUROC for logistic regression, random forest, sequential models in \texttt{Keras}, and \texttt{XGBoost} algorithms. To investigate the errors, we bootstrap the sample 100 times for each algorithms and estimate their uncertainties. To achieve high performance, we expect to have high TNR, low FPR, low FNR, high TPR, high accuracy, high precision, high recall, high F1 score, and high AUROC.  In general, the random forest, the sequential model in \texttt{Keras}, and the \texttt{XGBoost} algorithms provide higher TNR and lower FPR than the logistic regression algorithm, which is kind of a benchmark in binary classification.  The random forest can reach the highest precision, but \texttt{XGBoost} provides the lower FNR, higher TPR, higher recall, higher F1, and the highest AUROC, and still have high TNR, low FPR, and high precision. Considering the purpose to select enough AGN sources correctly, we chose \texttt{XGBoost} to identify AGNs and discuss the accuracy, precision, recall, F1 score and AUROC for different sample in the following. 

We perform grid-searching to test the performances of the input parameters with \texttt{XGBoost} as shown in Figure~\ref{xgb_roc_f1}. We find that performance does not improve significantly beyond 100 iterations, so it is reasonable that our training stop before 300 number round of iteration as mentioned in the previous subsection. In Figure~\ref{xgb_roc_f1}, brighter sample show better AUROC and F1 score for XAGNs and IRAGNs in \texttt{XGBoost}.  By considering both the sample size and the performance, we decided to use $r_{mag}<$23 (cModel photometry is adopted in this paper) as a cut for our subsample of 8,193 bright galaxies. 

We show the performances for our parent sample (32,204 objects without missing value of photometric redshift, HSC $girzy$, W1, and W2) in Table~\ref{tab_all}. We also show the performances of a bright sample (8,193 objects with $r_{mag}<$23) in Table~\ref{tab_sub}.
We list several choices of features: HSC only ($grizy$+photometric redshift), HSC+W12 ($grizy$+W1+W2+photometric redshift), HSC+WISE ($grizy$+W1+W2+W3+W4+photometric redshift), and WISE only (W1+W2+W3+W4+photometric redshift). 
In general, the performances of the feature choices can be ranked by: HSC+W12 $\sim$ HSC+WISE $>$ HSC only $>$ WISE only. It suggests that the additional infrared photometry (both HSC+W12 and HSC+WISE) is helpful for the classification. However, the performances can not be improved further if we include all WISE bands, perhaps because of the missing data in W3 and W4 bands, as well as the complexity to include upper limit in the learning process. Therefore, we adopted HSC+W12 for further tests in this paper. Moreover, pure optical information (HSC) can still identify AGNs, but it is not the case for using infrared information (WISE) only. 

Comparing Table~\ref{tab_all} and Table~\ref{tab_sub}, the trained machine performs better for the sample of the bright objects, especially for XAGN and IRAGN sample. Therefore, we focus on this sample to discuss the identification of different AGN types. We find that the performance of bright RAGN can not be improved , which is consistent with the grid-searching results in Figure~\ref{xgb_roc_f1}. 

In Table~\ref{tab_type}, we separated XAGN sample to 130 broad-line (type-1) and 241 non-broad-line (type-2) by their spectral type according to \citet{2016ApJ...819...62C,2016ApJ...817...34M}. We also test a classification of the XAGN sample according to their best-fitting SED templates and find similar results. 
We also separate IRAGN sample to 110 unobscured (type-1) and 277 obscured (type-2) by their best-fitting SED templates according to \citet{2017ApJS..233...19C}. As a result, we find that both XAGN (type-1) and IRAGN (type-1) have much better performance than XAGN (type-2) and IRAGN (type-2).

There are 197 XAGNs and IRAGNs (X+IRGNs) in common,  289 XAGNs and RAGNs (X+RGNs) in common, 180 IRAGNs and RAGNs (IR+RGNs) in common, as well as 114 AGNs belong to three AGN-selection methods (X+IR+RGNs).  We test the performance for AGN which belong to more than one sample as shown in Table~\ref{tab_common}.  The performance is slightly better but close to single AGN-selection method. 

Finally, we test the performance as a function of redshift, by breaking the sample into 3 redshift bins ($0.5<z<1.5$, $1.5<z<2.5$, and $2.5<z<3.5$), as shown in Table~\ref{tab_z}.
The performances are very similar at all redshift bins, as well as the whole sample. 
It suggests that redshift ranges do not seem to be a dominant factor to affect the performances, and \texttt{XGBoost} might be able to distinguish the sample with their redshift and photometry information.

\begin{table}[t]
\centering
\small
\setlength{\tabcolsep}{1.0pt}
\renewcommand{\arraystretch}{1.0} 
\begin{tabular}{cccccc}
\hline
\hline
type & ACC & P & R & F1 & AUROC  \\
\hline
\hline
\multicolumn{6}{c}{XGBoost: HSC only} \\ 
\hline
XAGN & 0.75$\pm$0.07 & 0.52$\pm$0.10 & 0.74$\pm$0.02 & 0.47$\pm$0.10 & 0.83$\pm$0.02 \\
IRAGN & 0.99$\pm$0.05 & 0.64$\pm$0.12 & 0.59$\pm$0.03 & 0.61$\pm$0.11 & 0.69$\pm$0.03 \\
RAGN & 0.85$\pm$0.05 & 0.59$\pm$0.05 & 0.65$\pm$0.01 & 0.60$\pm$0.05 & 0.81$\pm$0.02 \\
\hline
\multicolumn{6}{c}{XGBoost: HSC+W12} \\ 
\hline
XAGN & 0.98$\pm$0.05 & 0.63$\pm$0.08 & 0.60$\pm$0.02 & 0.61$\pm$0.08 & 0.82$\pm$0.02 \\
IRAGN & 0.99$\pm$0.02 & 0.70$\pm$0.08 & 0.60$\pm$0.03 & 0.63$\pm$0.06 & 0.73$\pm$0.03 \\
RAGN & 0.87$\pm$0.01 & 0.61$\pm$0.03 & 0.68$\pm$0.01 & 0.63$\pm$0.02 & 0.83$\pm$0.01 \\
\hline
\multicolumn{6}{c}{XGBoost: HSC+WISE} \\ 
\hline
XAGN & 0.98$\pm$0.03 & 0.64$\pm$0.07 & 0.60$\pm$0.02 & 0.61$\pm$0.06 & 0.80$\pm$0.02 \\
IRAGN & 0.99$\pm$0.02 & 0.65$\pm$0.08 & 0.59$\pm$0.03 & 0.62$\pm$0.06 & 0.75$\pm$0.03 \\
RAGN & 0.86$\pm$0.02 & 0.60$\pm$0.04 & 0.69$\pm$0.01 & 0.63$\pm$0.03 & 0.83$\pm$0.01 \\
\hline
\multicolumn{6}{c}{XGBoost: WISE only} \\ 
\hline
XAGN & 0.95$\pm$0.09 & 0.50$\pm$0.06 & 0.50$\pm$0.04 & 0.50$\pm$0.09 & 0.58$\pm$0.03 \\
IRAGN & 0.97$\pm$0.02 & 0.52$\pm$0.03 & 0.54$\pm$0.03 & 0.53$\pm$0.03 & 0.66$\pm$0.03 \\
RAGN & 0.60$\pm$0.12 & 0.53$\pm$0.05 & 0.59$\pm$0.05 & 0.46$\pm$0.09 & 0.63$\pm$0.04 \\
\hline
\hline
\end{tabular}
\caption{Numbers of accuracy, precision, recall, F1 score, and AUROC for XAGNs, IRAGNs, and RAGNs (All available objects; N=32,204). The uncertainties are derived by bootstrapping.}
\label{tab_all}
\end{table}

\begin{table}
\centering
\small
\setlength{\tabcolsep}{1.0pt}
\renewcommand{\arraystretch}{1.0} 
\begin{tabular}{cccccc}
\hline
\hline
type & ACC & P & R & F1 & AUROC \\
\hline
\hline
\multicolumn{6}{c}{XGBoost: HSC only} \\ 
\hline
XAGN & 0.94$\pm$0.04 & 0.64$\pm$0.10 & 0.62$\pm$0.03 & 0.63$\pm$0.08 & 0.78$\pm$0.03 \\
IRAGN & 0.98$\pm$0.01 & 0.64$\pm$0.07 & 0.66$\pm$0.04 & 0.65$\pm$0.05 & 0.74$\pm$0.04 \\
RAGN & 0.54$\pm$0.10 & 0.59$\pm$0.06 & 0.64$\pm$0.03 & 0.48$\pm$0.08 & 0.70$\pm$0.04 \\
\hline
\multicolumn{6}{c}{XGBoost: HSC+W12} \\ 
\hline
XAGN & 0.95$\pm$0.03 & 0.69$\pm$0.09 & 0.64$\pm$0.02 & 0.66$\pm$0.07 & 0.82$\pm$0.02 \\
IRAGN & 0.98$\pm$0.01 & 0.68$\pm$0.05 & 0.69$\pm$0.03 & 0.68$\pm$0.03 & 0.86$\pm$0.02 \\
RAGN & 0.81$\pm$0.06 & 0.62$\pm$0.04 & 0.65$\pm$0.02 & 0.63$\pm$0.05 & 0.77$\pm$0.02 \\
\hline
\multicolumn{6}{c}{XGBoost: HSC+WISE} \\ 
\hline
XAGN & 0.96$\pm$0.02 & 0.75$\pm$0.07 & 0.64$\pm$0.03 & 0.68$\pm$0.05 & 0.82$\pm$0.02 \\
IRAGN & 0.97$\pm$0.00 & 0.63$\pm$0.04 & 0.71$\pm$0.04 & 0.66$\pm$0.03 & 0.87$\pm$0.03 \\
RAGN & 0.79$\pm$0.06 & 0.62$\pm$0.05 & 0.68$\pm$0.02 & 0.64$\pm$0.06 & 0.78$\pm$0.02 \\
\hline
\multicolumn{6}{c}{XGBoost: WISE only} \\ 
\hline
XAGN & 0.93$\pm$0.07 & 0.55$\pm$0.08 & 0.54$\pm$0.03 & 0.55$\pm$0.08 & 0.65$\pm$0.03 \\
IRAGN & 0.96$\pm$0.01 & 0.54$\pm$0.05 & 0.57$\pm$0.04 & 0.55$\pm$0.03 & 0.75$\pm$0.04 \\
RAGN & 0.76$\pm$0.09 & 0.57$\pm$0.05 & 0.60$\pm$0.04 & 0.58$\pm$0.07 & 0.65$\pm$0.03 \\
\hline
\hline
\end{tabular}
\caption{Numbers of accuracy, precision, recall, F1 score, and AUROC for XAGNs, IRAGNs, and RAGNs ($r_{mag}<$23; N=8,193). The uncertainties are derived by bootstrapping.}
\label{tab_sub}
\end{table}

\begin{table}
\centering
\footnotesize
\setlength{\tabcolsep}{1.0pt}
\renewcommand{\arraystretch}{1.0} 
\begin{tabular}{cccccc}
\hline
\hline
type & ACC & P & R & F1 & AUROC \\
\hline
\multicolumn{6}{c}{XGBoost: HSC+W12} \\ 
\hline
XAGN & 0.98$\pm$0.05 & 0.63$\pm$0.08 & 0.60$\pm$0.02 & 0.61$\pm$0.08 & 0.82$\pm$0.02 \\
XAGN(type-1) & 0.99$\pm$0.00 & 0.78$\pm$0.04 & 0.75$\pm$0.05 & 0.76$\pm$0.04 & 0.95$\pm$0.02 \\
XAGN(type-2) & 0.99$\pm$0.00 & 0.54$\pm$0.04 & 0.52$\pm$0.03 & 0.53$\pm$0.03 & 0.78$\pm$0.02 \\
IRAGN & 0.99$\pm$0.02 & 0.70$\pm$0.08 & 0.60$\pm$0.03 & 0.63$\pm$0.06 & 0.73$\pm$0.03 \\
IRAGN(type-1) & 0.99$\pm$0.00 & 0.76$\pm$0.04 & 0.70$\pm$0.05 & 0.73$\pm$0.04 & 0.81$\pm$0.04 \\
IRAGN(type-2) & 0.99$\pm$0.00 & 0.58$\pm$0.52 & 0.75$\pm$0.03 & 0.53$\pm$0.03 & 0.66$\pm$0.03 \\
RAGN & 0.87$\pm$0.01 & 0.61$\pm$0.03 & 0.68$\pm$0.01 & 0.63$\pm$0.02 & 0.83$\pm$0.01 \\
\hline
\hline
\end{tabular}
\caption{Numbers of accuracy, precision, recall, F1 score,and AUROC for XAGNs (all+type1+type2), IRAGNs (all+type1+type2), RAGNs(all) for all available objects in the 32,204 parent sample. The uncertainties are derived by bootstrapping.}
\label{tab_type}
\end{table}

\begin{table}
\centering
\footnotesize
\setlength{\tabcolsep}{1.0pt}
\renewcommand{\arraystretch}{1.0} 
\begin{tabular}{cccccc}
\hline
\hline
type & ACC & P & R & F1 & AUROC \\
\hline
\hline
\multicolumn{6}{c}{XGBoost: HSC+W12} \\ 
\hline
XAGN & 0.98$\pm$0.05 & 0.63$\pm$0.08 & 0.60$\pm$0.02 & 0.61$\pm$0.08 & 0.82$\pm$0.02 \\
IRAGN & 0.99$\pm$0.02 & 0.70$\pm$0.08 & 0.60$\pm$0.03 & 0.63$\pm$0.06 & 0.73$\pm$0.03 \\
RAGN & 0.87$\pm$0.01 & 0.61$\pm$0.03 & 0.68$\pm$0.01 & 0.63$\pm$0.02 & 0.83$\pm$0.01 \\
\hline
X+IRAGN & 0.98$\pm$0.00 & 0.68$\pm$0.03 & 0.69$\pm$0.03 & 0.68$\pm$0.03 & 0.93$\pm$0.02 \\
X+RAGN & 0.98$\pm$0.00 & 0.68$\pm$0.03 & 0.69$\pm$0.03 & 0.68$\pm$0.03 & 0.92$\pm$0.02 \\
IR+RAGN & 0.99$\pm$0.00 & 0.66$\pm$0.04 & 0.64$\pm$0.04 & 0.65$\pm$0.03 & 0.86$\pm$0.03 \\
X+IR+RAGN & 0.99$\pm$0.00 & 0.68$\pm$0.05 & 0.63$\pm$0.05 & 0.65$\pm$0.05 & 0.92$\pm$0.03 \\
\hline
\hline
\end{tabular}
\caption{Numbers of accuracy, precision, recall, F1 score,and AUROC forXAGNs,  IRAGNs, RAGN, as well as AGNs in common: XAGNs+IRAGNs, XAGNs+RAGNs, IRAGNs+RAGNs, and XAGNs+IRAGNs+RAGNs in the 32,204 parent sample. The uncertainties are derived by bootstrapping.}
\label{tab_common}
\end{table}

\begin{table}
\centering
\footnotesize
\setlength{\tabcolsep}{1.0pt}
\renewcommand{\arraystretch}{1.0} 
\begin{tabular}{ccccccc}
\hline
\hline
type & number & ACC & P & R & F1 & AUROC \\
\hline
\hline
\multicolumn{7}{c}{XGBoost: HSC+W12 (All; N=32,204)} \\ 
\hline
XAGN & 625 & 0.98$\pm$0.05 & 0.63$\pm$0.08 & 0.60$\pm$0.02 & 0.61$\pm$0.08 & 0.82$\pm$0.02 \\
IRAGN & 404 & 0.99$\pm$0.02 & 0.70$\pm$0.08 & 0.60$\pm$0.03 & 0.63$\pm$0.06 & 0.73$\pm$0.03 \\
RAGN & 2225 & 0.87$\pm$0.01 & 0.61$\pm$0.03 & 0.68$\pm$0.01 & 0.63$\pm$0.02 & 0.83$\pm$0.01 \\
\hline
\multicolumn{7}{c}{XGBoost: HSC+W12 ($0.5<z<1.5$; N=21,342)} \\ 
\hline
XAGN & 485 & 0.98$\pm$0.00 & 0.64$\pm$0.04 & 0.60$\pm$0.02 & 0.62$\pm$0.02 & 0.82$\pm$0.02 \\
IRAGN & 254 & 0.99$\pm$0.02 & 0.69$\pm$0.07 & 0.60$\pm$0.03 & 0.63$\pm$0.05 & 0.73$\pm$0.04 \\
RAGN & 1640 & 0.86$\pm$0.02 & 0.60$\pm$0.03 & 0.67$\pm$0.01 & 0.62$\pm$0.03 & 0.82$\pm$0.01 \\
\hline
\multicolumn{7}{c}{XGBoost: HSC+W12 ($1.5<z<2.5$; N=2,005)} \\ 
\hline
XAGN & 39 & 0.98$\pm$0.00 & 0.63$\pm$0.08 & 0.59$\pm$0.07 & 0.61$\pm$0.06 & 0.81$\pm$0.07 \\
IRAGN & 46 & 0.99$\pm$0.01 & 0.67$\pm$0.07 & 0.61$\pm$0.05 & 0.63$\pm$0.05 & 0.72$\pm$0.06 \\
RAGN & 38 & 0.85$\pm$0.01 & 0.60$\pm$0.09 & 0.69$\pm$0.09 & 0.63$\pm$0.07 & 0.83$\pm$0.07 \\
\hline
\multicolumn{7}{c}{XGBoost: HSC+W12 ($2.5<z<3.5$; N=2,524)} \\ 
\hline
XAGN & 8 & 0.97$\pm$0.00 & 0.59$\pm$0.18 & 0.62$\pm$0.19 & 0.60$\pm$0.16 & 0.83$\pm$0.12 \\
IRAGN & 18 & 0.99$\pm$0.01 & 0.66$\pm$0.14 & 0.61$\pm$0.11 & 0.63$\pm$0.11 & 0.72$\pm$0.14 \\
RAGN & 12 & 0.88$\pm$0.00 & 0.60$\pm$0.19 & 0.65$\pm$0.16 & 0.62$\pm$0.15 & 0.83$\pm$0.18 \\
\hline
\hline
\end{tabular}
\caption{Numbers of sample size, accuracy, precision, recall, F1 score, and AUROC for XAGNs, IRAGNs, and RAGNs at different redshift bins. The uncertainties are derived by bootstrapping.}
\label{tab_z}
\end{table}

\subsection{Feature Importance}

\begin{figure*}
\centering
\includegraphics[width=0.32\textwidth]{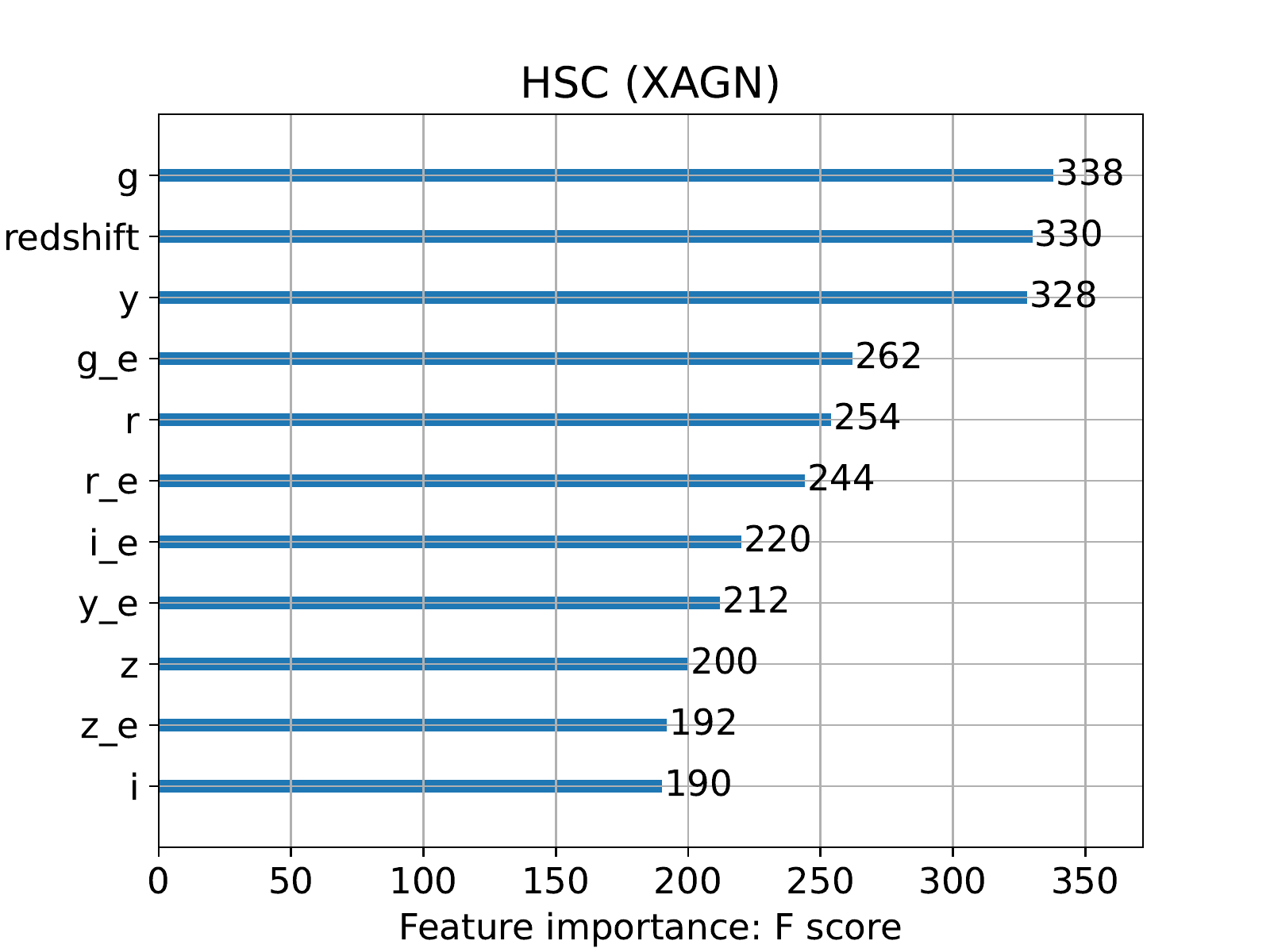} 
\includegraphics[width=0.32\textwidth]{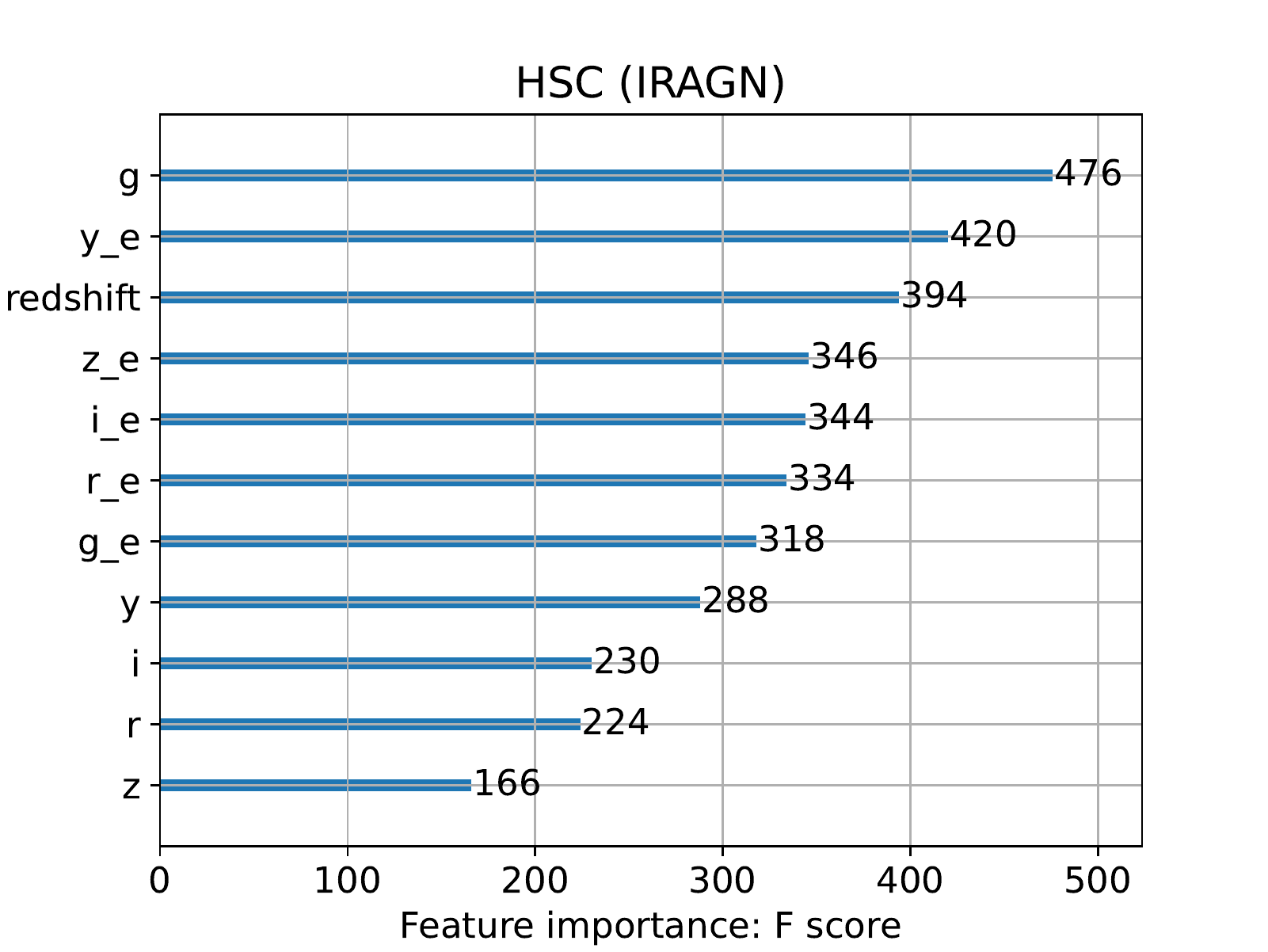} 
\includegraphics[width=0.32\textwidth]{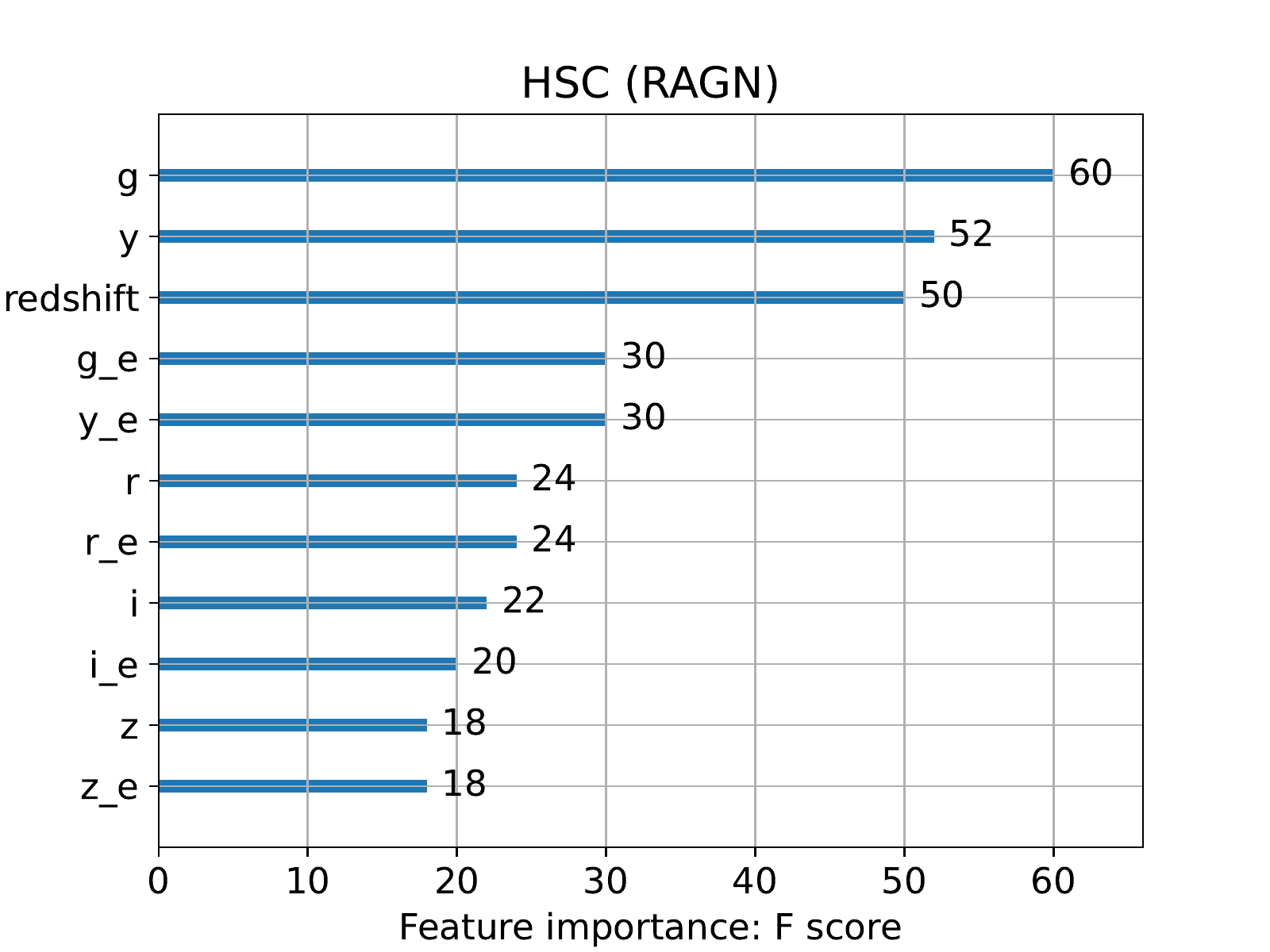} 
\includegraphics[width=0.32\textwidth]{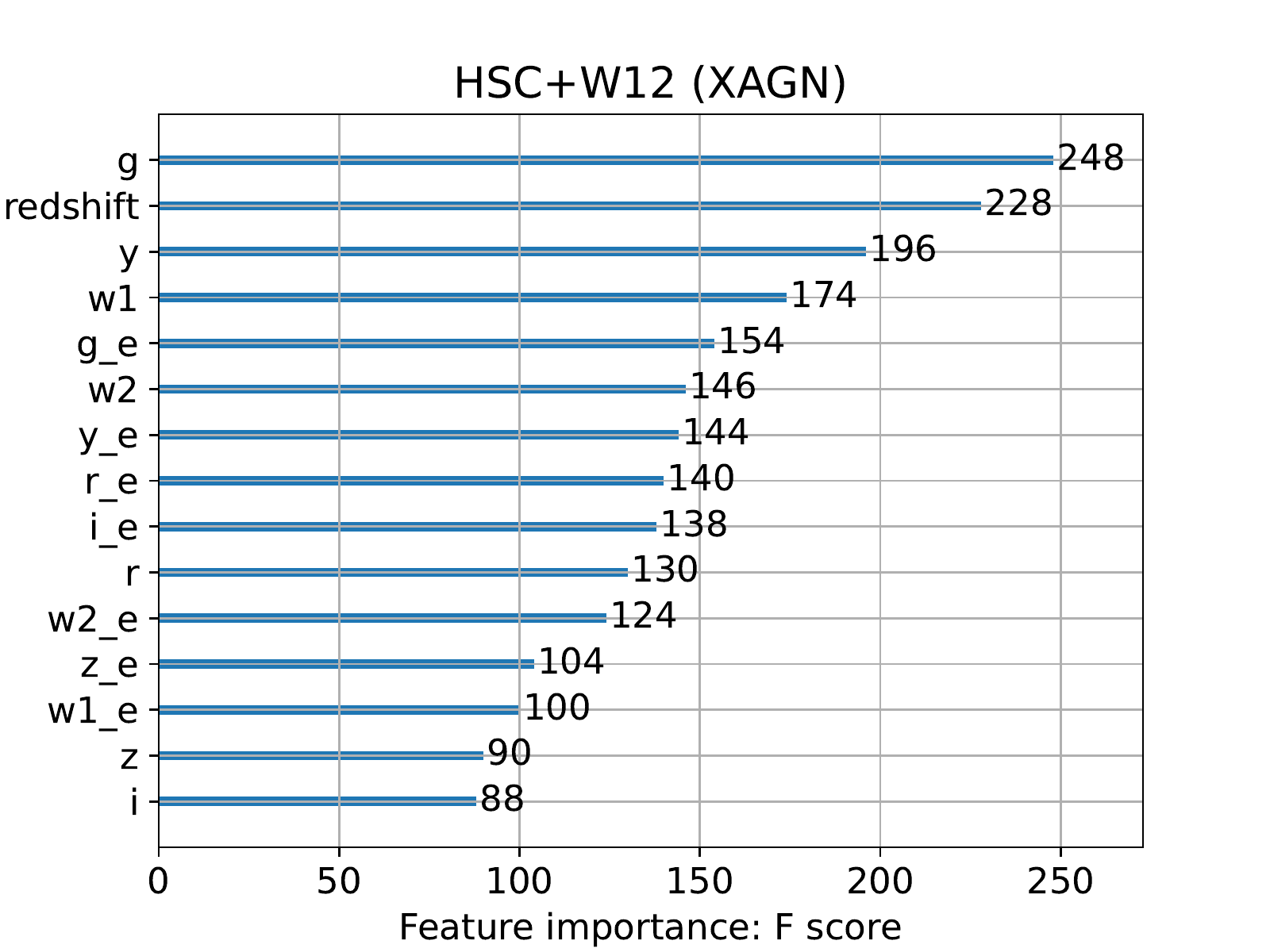} 
\includegraphics[width=0.32\textwidth]{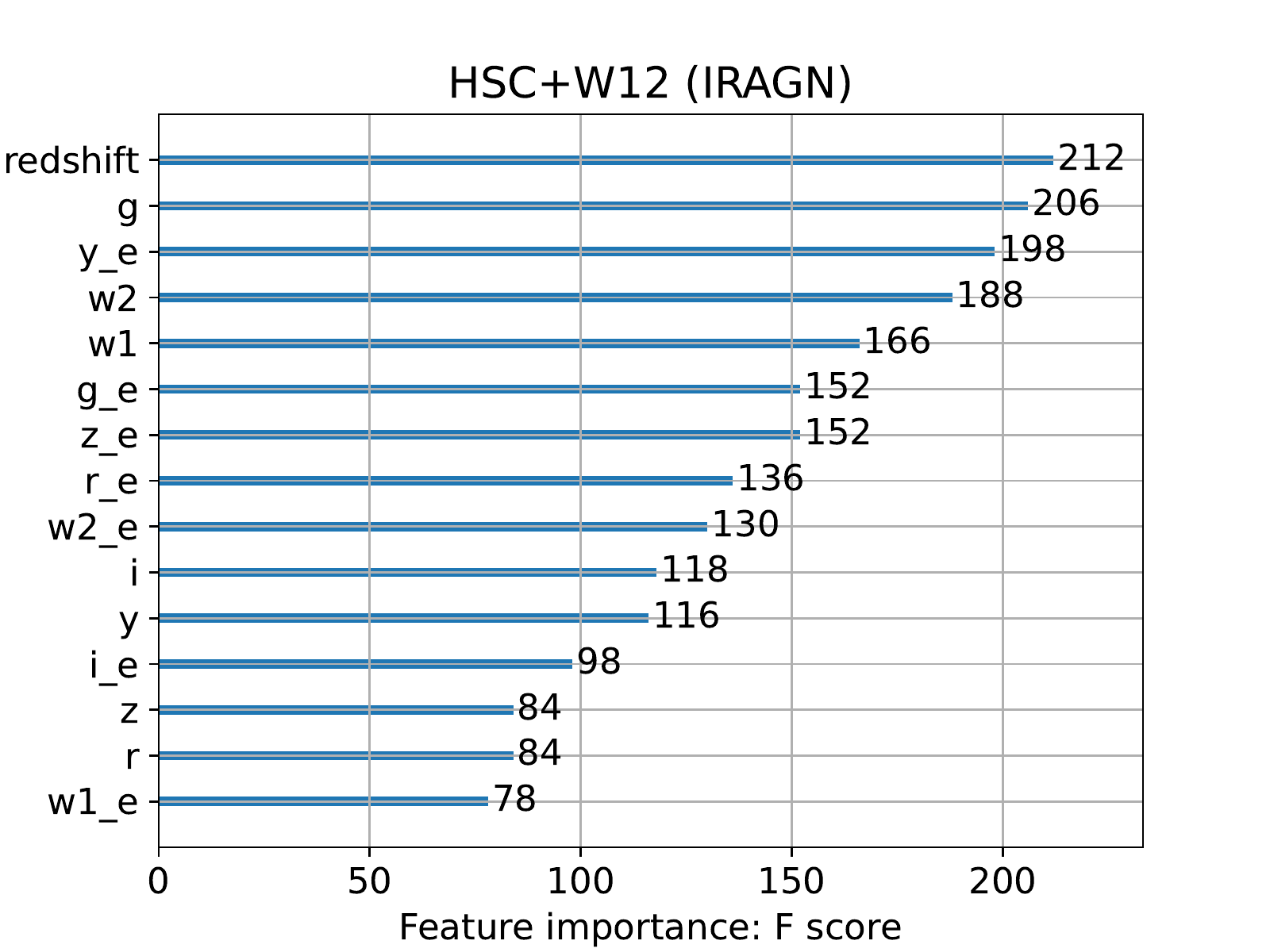} 
\includegraphics[width=0.32\textwidth]{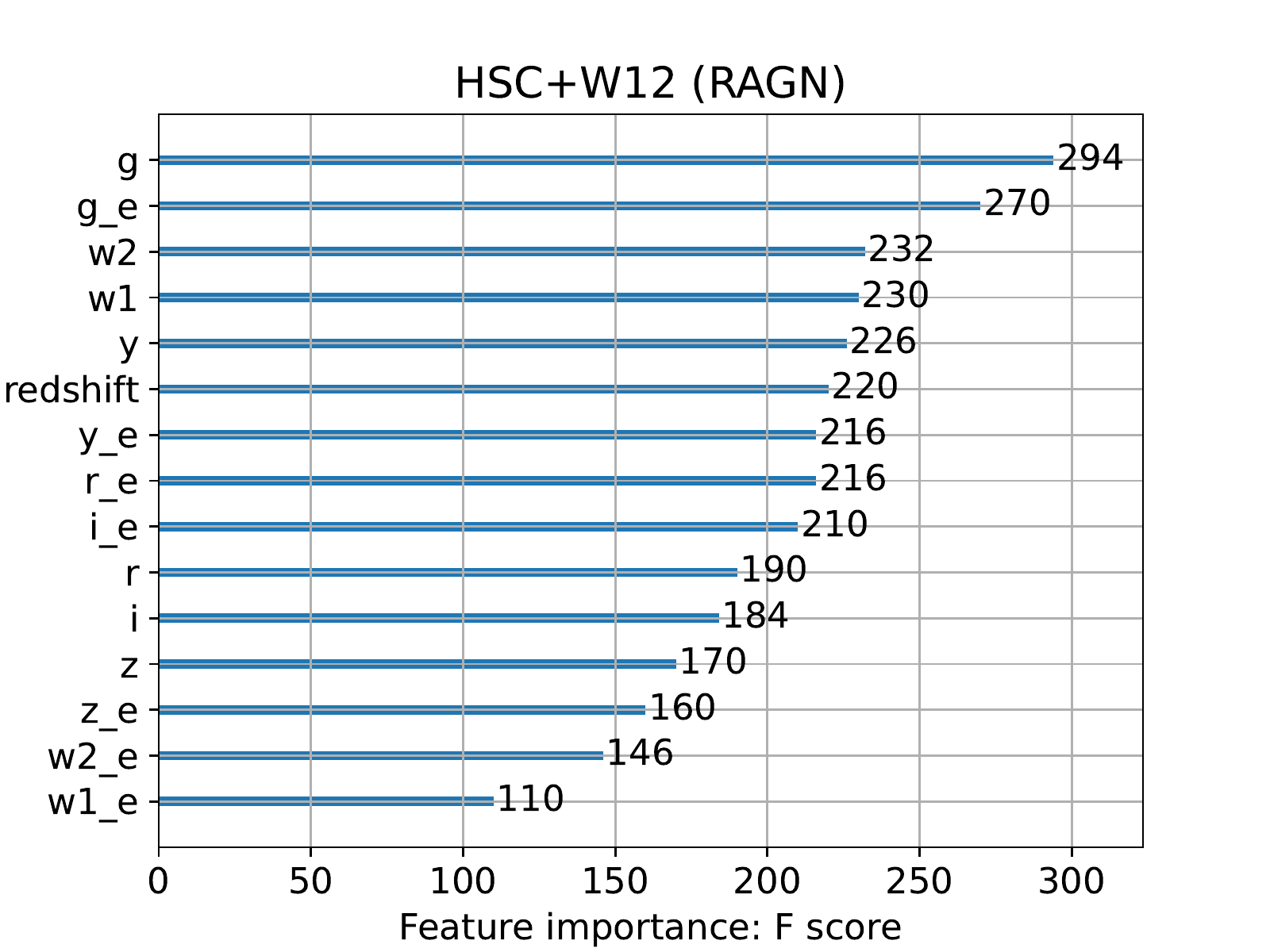} 
\includegraphics[width=0.32\textwidth]{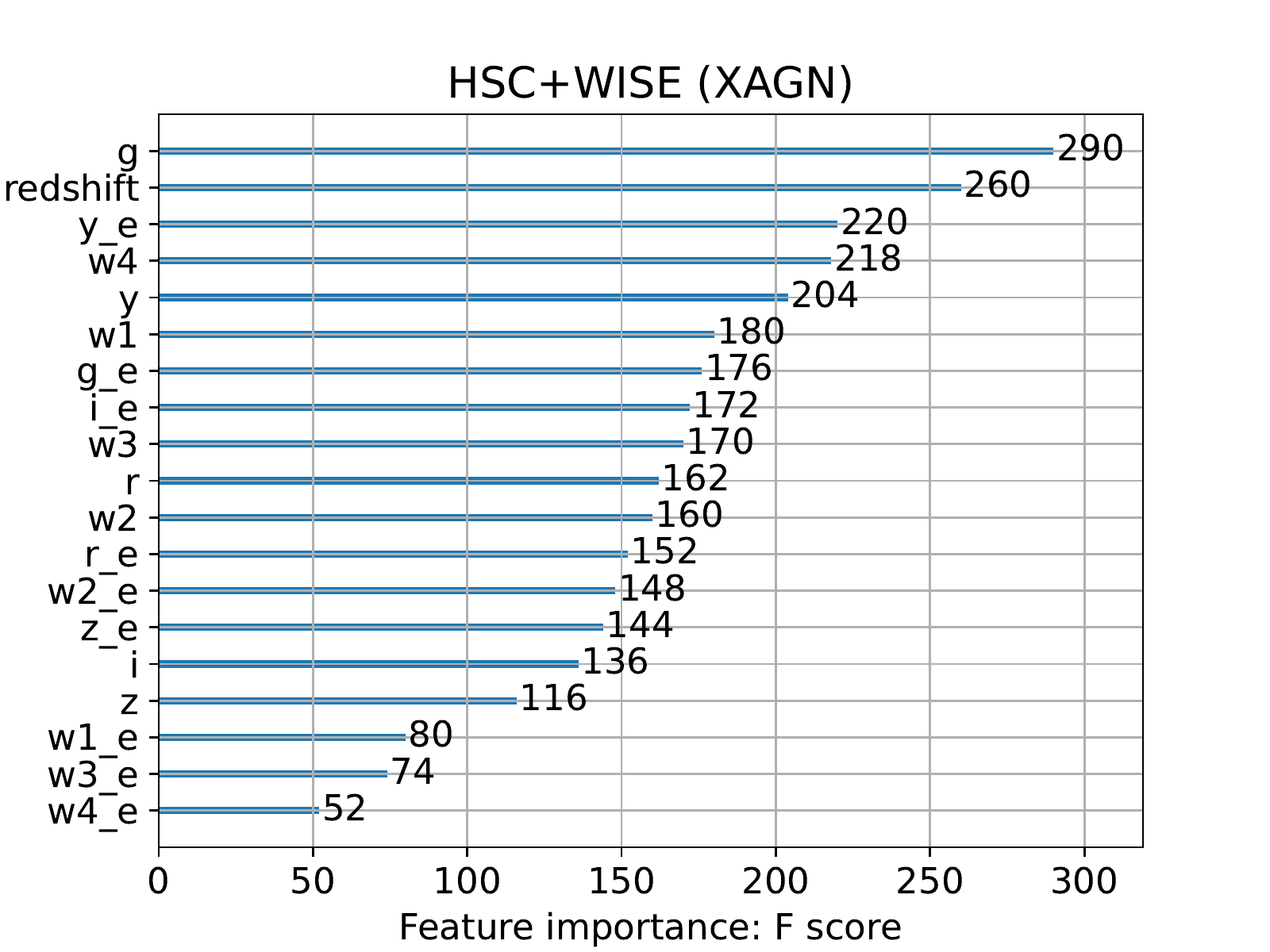} 
\includegraphics[width=0.32\textwidth]{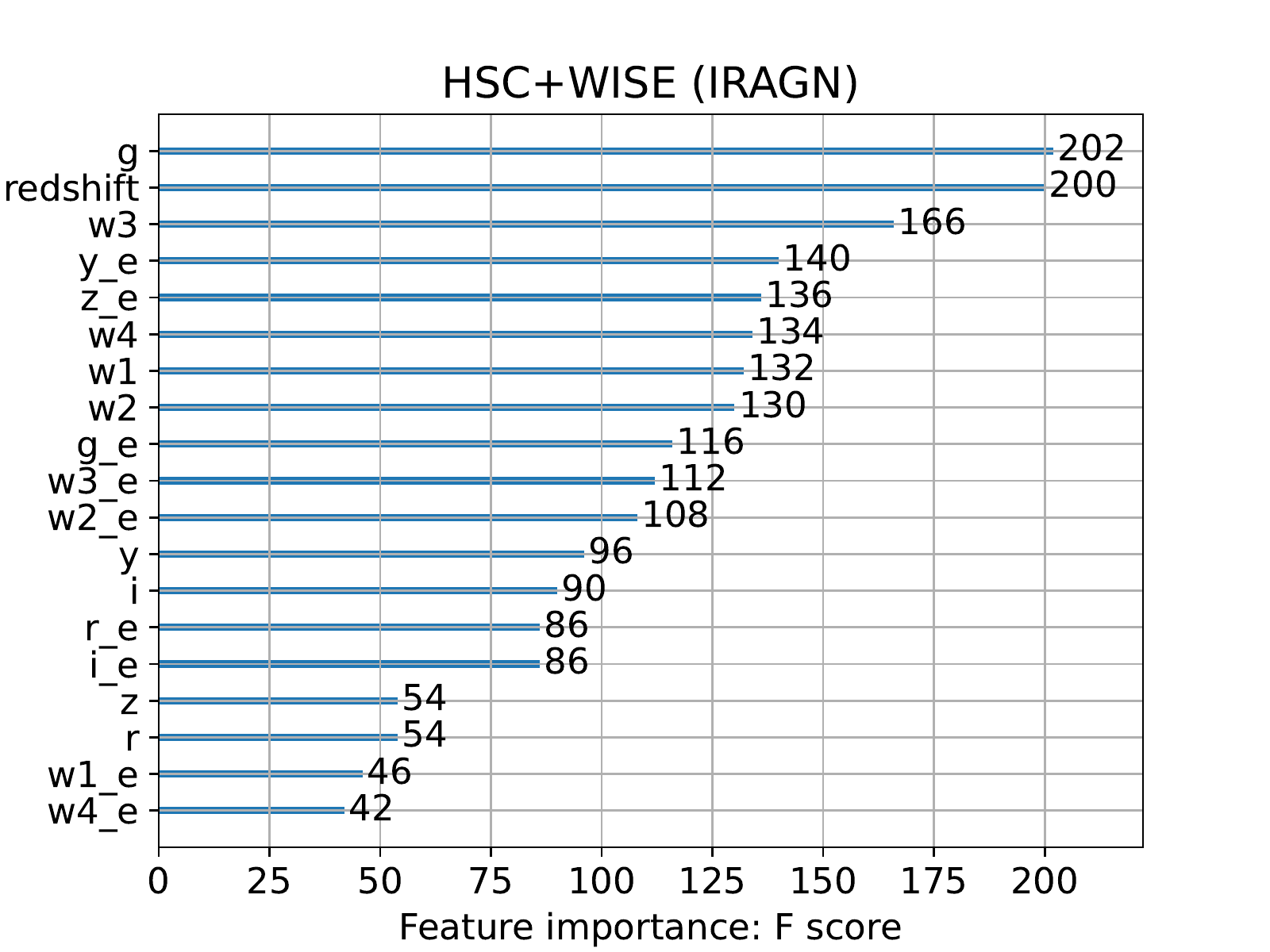} 
\includegraphics[width=0.32\textwidth]{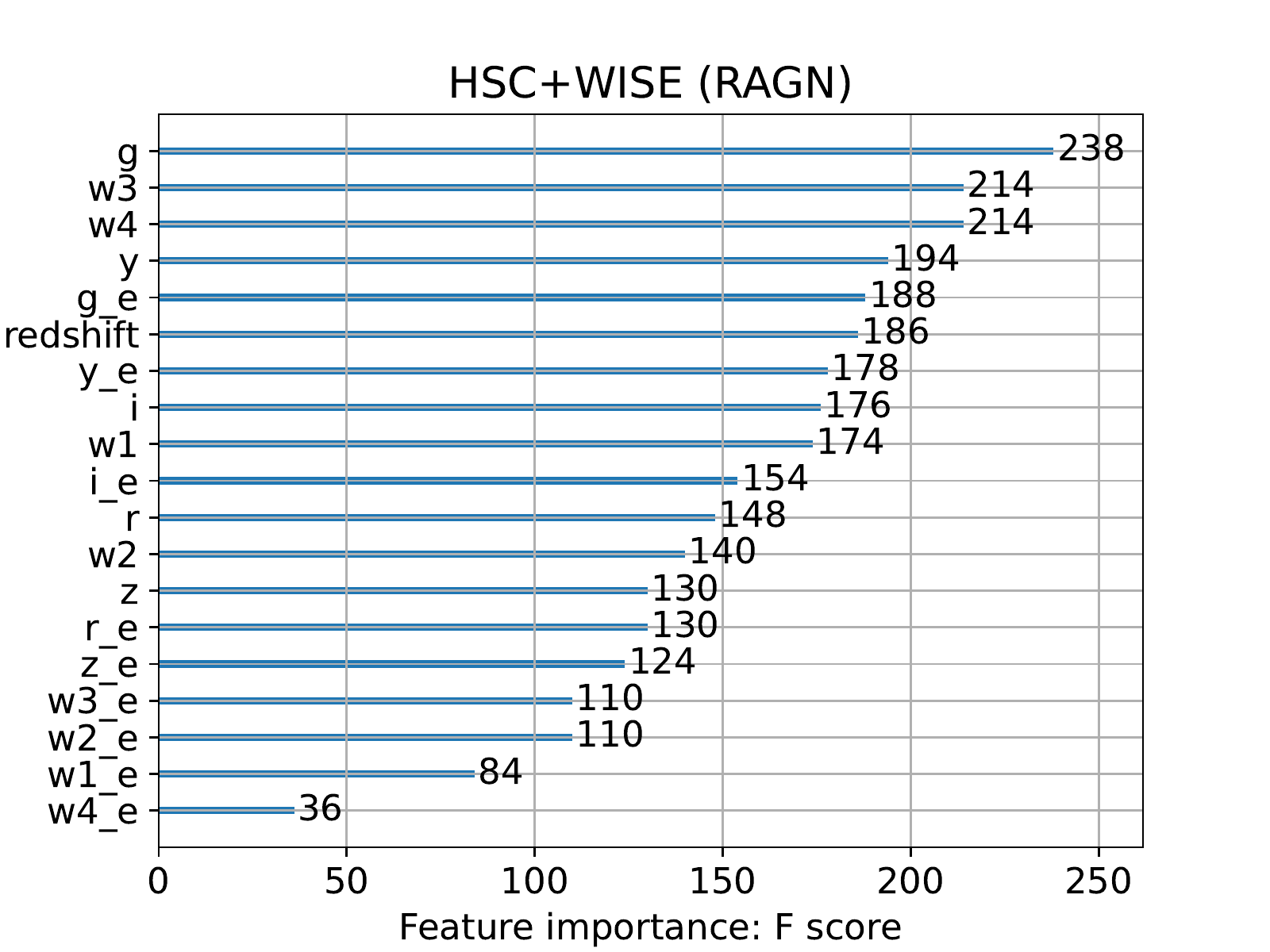} 
\includegraphics[width=0.32\textwidth]{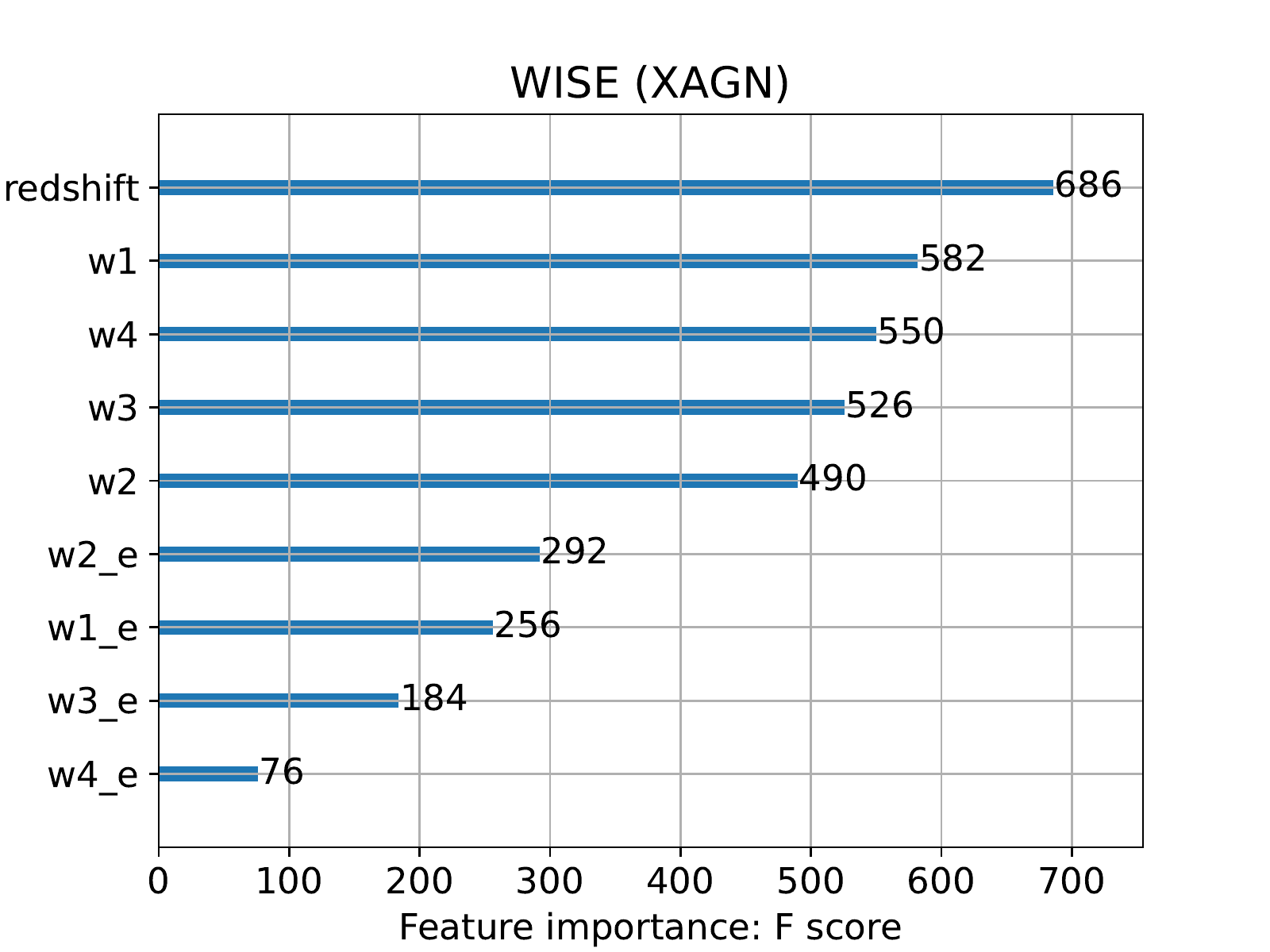} 
\includegraphics[width=0.32\textwidth]{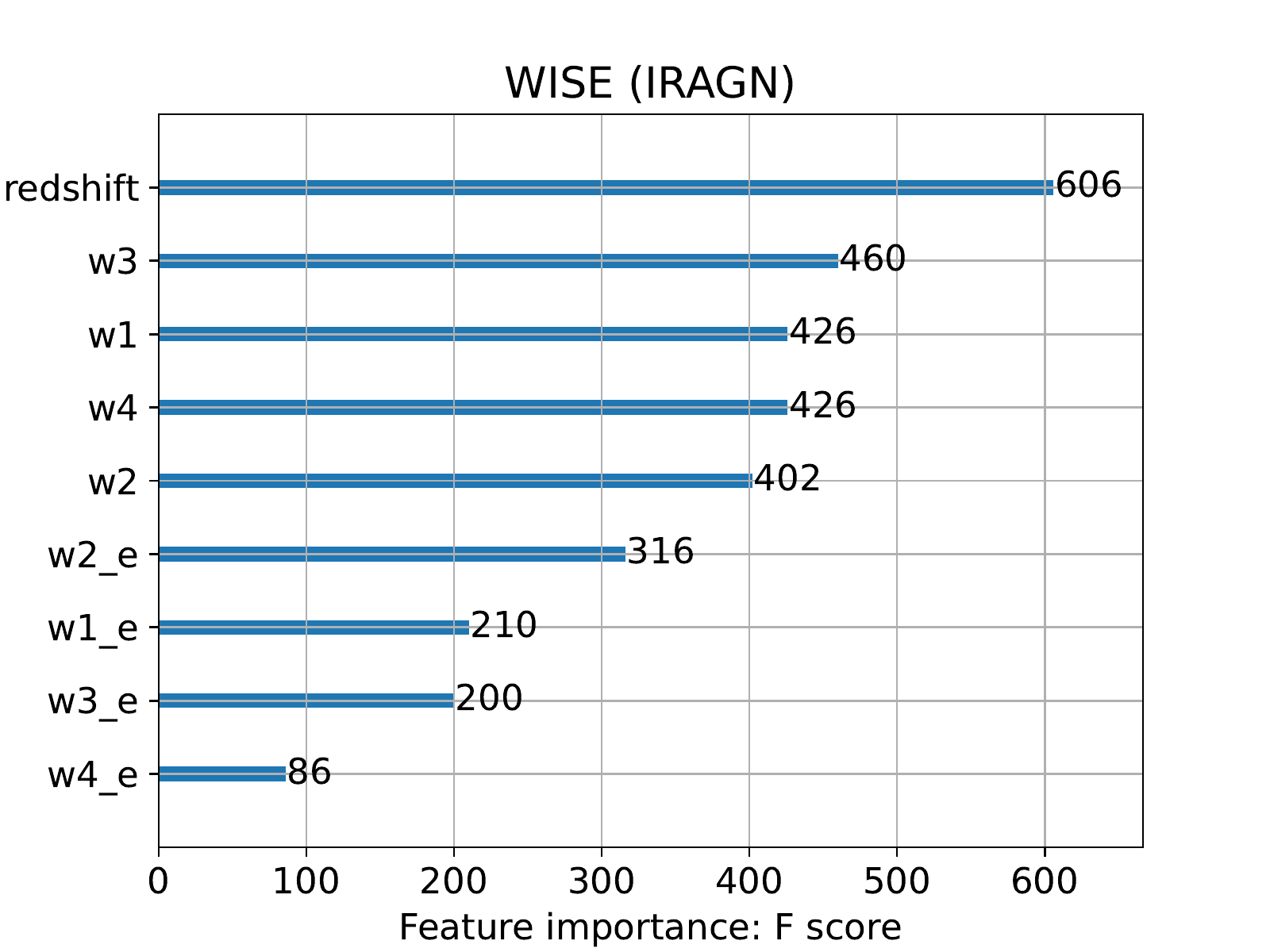} 
\includegraphics[width=0.32\textwidth]{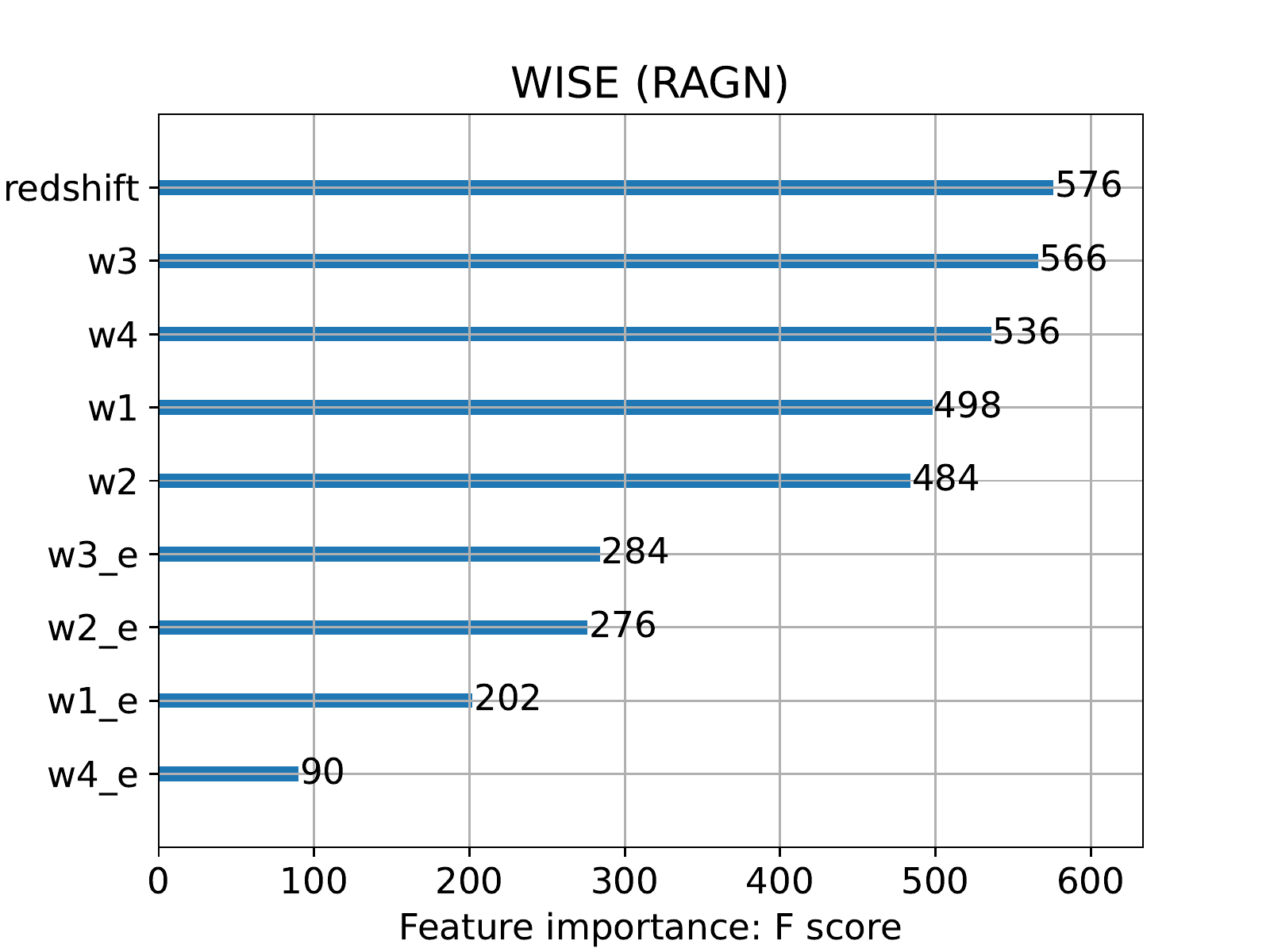} 
\caption{Feature importance of bright galaxies ($r_{mag}<$23; N=8,193) for XAGN (1st column), IRAGN (2nd column), and RAGN (3rd column).}
\label{xgb_importance}
\end{figure*}

We show the feature importance of \texttt{XGBoost} in Figure~\ref{xgb_importance} for our bright sample ($r_{mag}<$23; N=8,193), which indicates the importances of each input feature. 
For all the contributions derived from photometry, photometric redshift is the important feature for XAGNs (the 2nd, the 2nd, the 2nd, and the 1st important features for HSC, HSC+W12, HSC+WISE, and WISE sample, respectively), IRAGNs (the 3rd, the 1st, the 2nd, and the 1st important features for HSC, HSC+W12, HSC+WISE, and WISE sample, respectively), RAGNs (the 2nd, the 6th, the 6th, and 1st important features for HSC, HSC+W12, HSC+WISE, and WISE sample, respectively). The differences between types can be explained by their different selection approaches. 

If we only consider HSC photometry and the redshift in the first row (HSC only), $g$-band is the most important feature for all kinds of AGN. If we add W1 and W2 photometry as in the second row (HSC+W12), it is clear to see that they contribute to the identification for XAGNs (the 4th important feature), IRAGNs (the 4th important feature) and RAGNs (the 3rd important feature). If we also add W3 and W4 photometry as in the third row (HSC+WISE), W4 starts to be one of the important features. For identification with only WISE data (WISE only), photometry redshift is the dominant feature, and W1, W3, and W3 are the 2nd important feature for XAGNs, IRAGNs, RAGNs, respectively. 

In general, optical bands (e.g., $g$-band, $y$-band, and their uncertainties) as well as photometric redshift are the most important features, the former might be because of the data quality and the latter could be explained by that photometry redshift is already a combined information of the HSC photometry.  However, it is difficult to solely focus on one band measurement because all photometry are correlated.  If we remove the most important feature (e.g., $g$-band photometry), other correlated features (e.g., $y$-band, $r$-band, $i$-band, and $z$-band photometry) would become the most important one and the performances would not have significant decreasing. 
Moreover, if we remove the photometric redshift from the features, the ranking of the feature importance does not change much. To show the results with better performance, we keep the version with photometric redshift information. 

We find that WISE photometry can provide important information for the classification. However, the performance would be as good as all WISE bands if we only consider W1/W2. Figure~\ref{xgb_importance} shows that W3/W4 can dominate the features if we include them. It can be explained by that there are still missing values for W3/W4 in our subsample, and the sensitivities of W3/W4 (0.86/5.4 mJy) are much worse than W1/W2(0.068/0.098 mJy). 

\begin{table}
\center
\scriptsize
\setlength{\tabcolsep}{1.0pt}
\renewcommand{\arraystretch}{1.0} 
\begin{tabularx}{1.0\linewidth}{lllll}
\hline
\hline
Bytes & Format & Units &  Label & Explanations\\
\hline
 1-17 & I17 & --- & objid & objid in HSC-SSP PDR2\\
19-27 & F3.5 & deg & ra & right ascension in HSC-SSP PDR2\\
29-35 & F3.5 & deg & dec & declination in HSC-SSP PDR2\\
37-37 & I1 & --- & XAGN & 1: XAGNs predicted by ML\\
39-39 & I1 & --- & IRAGN & 1: IAGNs predicted by ML\\
41-41 & I1 & --- & RAGN & 1: RAGNs predicted by ML\\
43-49 &  F1.5 &  --- & XAGNp & Probability of X-ray selected AGNs\\
51-57 &  F1.5 &  --- & IRAGNp & Probability of Infrared selected AGNs\\
59-65  & F1.5 &  --- & RAGNp & Probability of Radio selected AGNs\\
\end{tabularx}
\begin{tabularx}{1.0\linewidth}{ccccccccc}
\hline
43153636661927346 & 149.45994 & 1.73713 & 0 & 0 & 1 & 0.47351 & 0.48839 & 0.84884\\
43153636661946779 & 149.45624 & 1.71012 & 0 & 0 & 0 & 0.02193 & 0.06230 & 0.41638\\
43153636661947106 & 149.49619 & 1.71407 & 0 & 0 & 1 & 0.20267 & 0.12076 & 0.59937\\
43153636661948529 & 149.42880 & 1.73570 & 0 & 0 & 0 & 0.05203 & 0.12532 & 0.19787\\
: & : & : & : & : & : & : & : & :\\
\hline
\hline
\end{tabularx}
\caption{AGN candidates in the HSC-Wide region for 112,609 objects. The input parameters are photometric redshift, $ugriz$, W1, W2 bands. The training data is the 32,204 parent sample.  We apply XGBoost as described in Section~\ref{sec3}. The objects are predicted as AGNs when the probabilities are larger than 0.5. }
\tablecomments{Table~\ref{tab_cat} is published in its entirety in the machine-readable format. A portion is shown here for guidance regarding its form and content.}
\label{tab_cat}
\end{table}

\section{Discussion} \label{sec4}

\subsection{Can optical to infrared data identify AGN Hosts?}

We use \texttt{scikit-learn}, \texttt{Keras}, and \texttt{XGBoost} packages to identify AGN hosts from galaxy sample. We find that \texttt{XGBoost} can provide better performances (e.g., TNR, FPR, FNR, TPR, accuracy, precision, recall, F1 score, and ROC), perhaps because of the limit of the algorithms, the choice of our input parameters, and the characteristic of our data. 
In order to avoid the missing value problem due to non-detected data in the catalog, we consider 32,204  galaxies with available photometric redshifts, as well as detections in HSC $grizy$ bands, W1, and W2. In fact, \texttt{XGBoost} can also deal with missing value and perform well as shown in our HSC+WISE feature choice. 
In order to avoid imbalance problem between AGNs and GALs, we use over-sampling technique. For this kind of data, accuracy (AGN and non-AGN which are classified correctly over all sources) would be mainly determined by the non-AGN sample. Therefore, we have to focus on the F1 (harmonic mean of the precision and the recall) and the AUROC (TP versus FP) values. 

In general, optical HSC information with near-infrared information (HSC+W12 or HSC+WISE) show the best performance (F1 score $>$ 0.60; AUROC $>$ 0.70) among all feature combinations as shown in Table~\ref{tab_all} and Table~\ref{tab_sub}. Pure optical (HSC only) or pure infrared (WISE only) may still provide some information, but it seems better to combine them. Both HSC+W12 and HSC+WISE show good performance (F1 score $>$ 0.65; AUROC $>$ 0.75) for our bright sample. This is consistent with the performance of traditional machine learning and deep learning algorithms for galaxy morphology classification by \citep{2020A&C....3000334B}.
Moreover, pure infrared (WISE only) performance can be improved for the bright sample, which shows that the optical bands are deeper than the infrared bands. 
Besides, photometric redshift can be one of the important features and increase the performance by about 2\%. It can be explained by that the photometric redshift is derived by the optical photometry, so pure photometry information is enough to provide good performance. According to the results of feature importance, we notice that the input parameters are highly correlated. If we remove the most important feature (e.g., photometric redshift or $g$-band), other correlated features would become the most important feature and the performances would have no significant changes.  Nevertheless, we show that machine learning technique can have good performances to identify AGN hosts with optical to infrared data.

We apply the same machine-learning method to objects in the HSC-Wide layer for HSC-SSP Public Release Data 2 \citep[PDR2; ][]{2019PASJ...71..114A}. Among the 3,238,247 objects, there are 112,609 of them with available photometric redshift, HSC $grizy$, W1, and W2 bands. The criterion is similar to the 32,204 parent sample as described  in Section~\ref{sec2}, and the sample size is around 3.5 times larger. As described in Section~\ref{sec33} and Table~\ref{tab_tfpnr}, the output catalog can reach a high accuracy and good performance (F1 score $>$ 0.60; AUROC $>$ 0.70) by XGBoost. 
According to the precision and recall scores in Table~\ref{tab_tfpnr}, the purity is larger than 60\% and the completeness is larger than 60\%. Higher purity and completeness can be achieved by subsamples as discussed in Section~\ref{sec42}, but we keep the complete sample here.
According to our prediction, there are 23,157 XAGN,  9,541 IRAGN, and 38,876 RAGN candidates among the 112,609 objects.  
The predicted data provides dozens of time larger sample than the training data, which requires X-ray, infrared, or radio observations as described in Section~\ref{sec2}. 
Table~\ref{tab_cat} shows the predicted catalog for AGN candidates in the entire HSC-Wide region.
The catalog provides the probability (\texttt{XAGNp}, \texttt{IRAGNp}, and \texttt{RAGNp}) of being an AGN with each selection method as well as the AGN flag (\texttt{XAGN}, \texttt{IRAGN}, and \texttt{RAGN}) when the probability is larger than 0.5.

\subsection{Can we classify different kinds of AGNs?} \label{sec42}

The performance is high (F1 score $>$ 0.65; AUROC $>$ 0.80) for bright XAGN and IRAGN host galaxies as shown in Figure~\ref{xgb_roc_f1} and Table~\ref{tab_sub}. However, the performance of bright RAGN host galaxies can not be improved.  As shown in Figure~\ref{xgb_roc_f1}, a sample with bright galaxies show slightly worse AUROC and only slightly better F1 score for RAGN host galaxies.  To avoid imbalance between AGN types, we test the data with both over-sampling and under-sampling techniques, and find that the performance from XAGN and IRAGN sample is still better than RAGN sample. This might be explained by the sample selection of different kinds of AGN. For instance, IRAGNs are difficult to be selected if they are not bright enough in the infrared part, which can be highly correlated with the optical photometry. Because the RAGN sample size is about 3-5 times larger than XAGN and IRAGN sample, it may suggest that the intermediate bright AGNs with $>$ 3-$\sigma$ radio excess in $\log(L_{1.4GHz}/SFR_{IR}$) require more optical, IR, or radio observations to identify them.  Besides,  AGNs selected by multiple methods show slightly better performance than single methods as shown in Table~\ref{tab_common}.  It might because these are objects with significant AGN features.  However, the results are very close so larger sample would be needed to investigate the differences.

For both type-1 (broad-line) XAGNs and type-1 (unobscured) IRAGNs, the performance is very good (F1 score $>$ 0.70; AUROC $>$ 0.80) by using optical to infrared information as shown in Table~\ref{tab_type}. It can be explained by the point source like features of type-1 AGNs.  We test structural parameters of type-1 and type-2 AGNs by using GAFLIT \citep{2010AJ....139.2097P} results for HSC/ACS imaging in \citet{2017ApJS..233...19C}, which used single S\'ersic profile fitting with HSC/ACS imaging.  About 60\% of the AGN sample can be fitted with single S\'ersic profile. We find that the fitting radii of type-1 AGNs are significant smaller than that of type-2 AGNs (the significant probability by Kolmogorov-Smirnov test is smaller than 5\%).  And the S\'ersic indices of type-1 AGNs are marginally larger than that of type-2 AGNs. This suggests that the point source like feature of type-1 AGNs are distinguishable.  Moreover, the structural parameters of non-AGNs are also significant different from both type-1 and type-2 AGNs, but a detailed matched mass sample should be investigated. To avoid sample selection constraint by structural measurement,  we do not include HSC morphology as input parameters. A detailed studies for HSC imaging could be also explored by more sophisticated machine-learning techniques.  In this work, we focus on HSC photometry and show that  we are able to classify type-1 AGNs by our technique with very high performance.  On the other side, the performance is still fine  (F1 score $>$ 0.50; AUROC $>$ 0.60) for type-2 (non-broad-line) XAGNs and type-2 (obscured) IRAGNs.  In Table~\ref{tab_z}, we separate galaxies to different redshift bins, which show that redshift range is not a dominant factor to identify AGN host galaxies by machine learning. This is consistent with \citep{2018A&A...611A..97P} that uses CNN to classify and predict the photometric redshift of quasars in SDSS Stripe 82 via using light curves.  For example, XAGNs show better performance at $1.5<z<2.5$, but not RAGNs. It might be explained by the identification of the original AGN sample.  The sizes of our various subsample are not very large, especially for bright objects and individual bins.  We estimate uncertainties by bootstrapping for the whole learning process and it seem reasonable for most cases. Nevertheless, larger AGN sample in future observations would be able to constrain the subsamples better.

The main bands used are in the optical, so we also consider optical selected AGNs. We match our catalog with the optical spectral catalog in the Sloan Digital Sky Survey Data Release 16 \citep[SDSS-DR16;][]{2020ApJS..249....3A}. The classifications (\texttt{CLASS} and \texttt{SUBCLASS}) which are labeled as \texttt{QSO}, \texttt{AGN}, \texttt{BROADLINE}, and \texttt{AGN BROADLINE} in SDSS are adopted as optical selected AGNs. The sample size is quite small (38 AGNs out of 231 matched objects), but the performance is very good (F1 score = 0.86$\pm$0.05 ; AUROC = 0.93$\pm$0.06). It suggests that our method can also apply to optical selected AGNs, and a larger sample of training data are required to provide a general information. Besides, we test AGNs which are labeled by broad-band SED fitting in \citet{2017A&A...602A...3D} and \citet{2017ApJS..233...19C}. The performance is also fine (F1 score $>$ 0.60 ; AUROC $>$ 0.65). In this work, we focus on XAGNs, IRAGNs, and RAGNs. Nevertheless, machine-learning technique can also apply to other AGN selection methods in the future. 

We do a sanity check by validating data outside the COSMOS field in the HSC-Wide layer: XXL AGN sample \citep{2016MNRAS.457..110M}, SWIRE IR AGN sample \citep{2003PASP..115..897L}, and RAGN sample in \citep{2012MNRAS.421.1569B}. Though the techniques of AGN selection are slightly different, the performances (0.70 $>$ F1 score $>$ 0.50; 0.80 $>$ AUROC $>$ 0.60) show that our algorithms seem to work fine. 

In general, machine learning techniques are able to classify different kinds of AGNs, but the performances would depend on the AGN selections, the AGN types, the magnitude choices, and the sample size.  In the future, it would be also helpful for us to investigate other physical properties of AGNs and their host galaxies. 


\section{Summary} \label{sec5}
In this paper, we have identified XAGNs, IRAGNs, RAGNs, and GALs by machine learning techniques. Our main findings are as follows.
\begin{enumerate}
\item The HSC (optical) information with WISE band-1 and WISE band-2 (near-infrared) information perform well (F1 score $>$ 0.60; AUROC $>$ 0.70) to identify AGN hosts.
\item The performance is good (F1 score $>$ 0.65; AUROC $>$ 0.75) for bright XAGN and IRAGN host galaxies.
\item For both type-1 (broad-line) XAGNs and type-1 (unobscured) IRAGNs, the performance is very good (F1 score $>$ 0.70; AUROC $>$ 0.80).
\item These results can apply to the five-band data from the wide regions of the HSC survey, and future all-sky surveys. 
\end{enumerate}

\acknowledgments

We thank the referee as well as Chien-Hsiu Lee for helpful comments and discussions.
YYC acknowledge financial support from the Ministry of Science and Technology of Taiwan grant (109-2112-M-005 -003 -MY3, 108-2112-M-001-014-, and 105-2112-M-001-029-MY3). 
We gratefully acknowledge the contributions of the entire COSMOS collaboration.\\

The Hyper Suprime-Cam (HSC) collaboration includes the astronomical communities of Japan and Taiwan, and Princeton University. The HSC instrumentation and software were developed by the National Astronomical Observatory of Japan (NAOJ), the Kavli Institute for the Physics and Mathematics of the Universe (Kavli IPMU), the University of Tokyo, the High Energy Accelerator Research Organization (KEK), the Academia Sinica Institute for Astronomy and Astrophysics in Taiwan (ASIAA), and Princeton University. Funding was contributed by the FIRST program from Japanese Cabinet Office, the Ministry of Education, Culture, Sports, Science and Technology (MEXT), the Japan Society for the Promotion of Science (JSPS), Japan Science and Technology Agency (JST), the Toray Science Foundation, NAOJ, Kavli IPMU, KEK, ASIAA, and Princeton University. 
This paper makes use of software developed for the Large Synoptic Survey Telescope. We thank the LSST Project for making their code available as free software at  http://dm.lsst.org
This paper is based S19a on data collected at the Subaru Telescope and retrieved from the HSC data archive system, which is operated by Subaru Telescope and Astronomy Data Center at National Astronomical Observatory of Japan. Data analysis was in part carried out with the cooperation of Center for Computational Astrophysics, National Astronomical Observatory of Japan.

This publication makes use of data products from the Wide-field Infrared Survey Explorer, which is a joint project of the University of California, Los Angeles, and the Jet Propulsion Laboratory/California Institute of Technology, and NEOWISE, which is a project of the Jet Propulsion Laboratory/California Institute of Technology. WISE and NEOWISE are funded by the National Aeronautics and Space Administration.

%







\bibliographystyle{aasjournal}



\end{document}